\documentclass[12pt,letterpaper]{JHEP3}
\pdfoutput=1
\usepackage{amsmath,amssymb,scalefnt}
\usepackage{graphicx}
\def\al{\alpha} 
\def\ga{\gamma}

\def\th{\theta}

\def\la{\lambda}

\def\si{\sigma}

\def\om{\omega}
\def\De{\Delta}

\def\La{\Lambda}

\def\Om{\Omega}

\def\barm{\bar{m}}
\def\tilh{\tilde{h}}

\newcommand{\ben}{\begin{equation}}
\newcommand{\een}{\end{equation}}
\newcommand{\bea}{\begin{eqnarray}}
\newcommand{\eea}{\end{eqnarray}}
\newcommand{\ba}{\begin{array}}
\newcommand{\ea}{\end{array}}
\newcommand{\bit}{\begin{itemize}}
\newcommand{\eit}{\end{itemize}}

\def\math{\mathsurround 0pt}
\def\oversim#1#2{\lower.5pt\vbox{\baselineskip0pt \lineskip-.5pt
 \ialign{$\math#1\hfil##\hfil$\crcr#2\crcr{\scriptstyle\sim}\crcr}}}

\def\half{\frac{1}{2}}

\def\Tr{\mathop{\rm Tr}\nolimits}

\newcommand{\vev}[1]{\left\langle#1\right\rangle}

\newcommand{\tenfun}{B}
\newcommand{\powspec}{\mathcal{P}}

\newcommand{\mpl}{m_\mathrm{P}}

\newcommand{\eqn}[1]{Eq.~(\ref{#1})}
\newcommand{\eqns}[2]{Eqs.~(\ref{#1}),(\ref{#2})}

\newcommand{\reference}[1]{Ref.~\cite{#1}}
\newcommand{\references}[2]{Refs.~\cite{#1,#2}}
\newcommand{\referencesb}[3]{Refs.~\cite{#1,#2,#3}}

\newcommand{\abbrev}{\scalefont{.9}}

\newcommand{\sm}{{\abbrev SM}}
\newcommand{\rg}{{\abbrev RG}}

\newcommand{\cp}{{\abbrev CP}}
\newcommand{\lsp}{{\abbrev LSP}}

\newcommand{\mssm}{{\abbrev MSSM}}
\newcommand{\nmssm}{{\abbrev NMSSM}}

\newcommand{\msugra}{{\abbrev MSUGRA}}
\newcommand{\cmssm}{{\abbrev CMSSM}}
\newcommand{\amsb}{{\abbrev AMSB}}
\newcommand{\mamsb}{{\abbrev mAMSB}}
\newcommand{\samsb}{{\abbrev sAMSB}}
\newcommand{\RG}{{\abbrev RG}}

\def\ee{\end{equation}}

\def\frak#1#2{{\textstyle{\frac{#1}{#2}}}}
\def\half{\textstyle{\frac{1}{2}}}

\def\btil{\tilde b} 
 
\def\dtil{\tilde d} 
\def\etil{\tilde e}

\def\ttil{\tilde t}
\def\util{\tilde u}

\def\tautil{\tilde \tau}

\def\nutil{\tilde \nu}

\def\GeV{\hbox{GeV}}
\def\TeV{\hbox{TeV}}

\def\vev#1{\mathopen\langle #1\mathclose\rangle }
\def\nn{\nonumber\\}
\def\DRED{\ifmmode{{\rm DRED}} \else{{DRED}} \fi}
\def\DREDD{\ifmmode{{\rm DRED}'} \else{${\rm DRED}'$} \fi}
\def\NSVZ{\ifmmode{{\rm NSVZ}} \else{{NSVZ}} \fi}

\def\ga{\gamma}

\def \la{\lambda}
\def \La{\Lambda}
\def \th{\theta}
\def\sic{supersymmetric}

\def\sy{supersymmetry}
\def\sic{supersymmetric}

\def\phib{\overline{\phi}}
\def\Phib{\overline\Phi}

\author{Mark~Hindmarsh$^{1,3}$, 
D.~R.~Timothy~Jones$^{2}$\\
$^{1}$Dept.\ of Physics and Astronomy, University of Sussex, 
Brighton BN1 9QH, U.K.\\
$^{2}$Dept. of Mathematical Sciences,
University of Liverpool, Liverpool L69 3BX, U.K.\\
$^{3}$Helsinki Institute of Physics, P.O.\ Box 64, 00014 Helsinki University, Finland
}

\title{Strictly Anomaly Mediated Supersymmetry Breaking}

\abstract{
We consider an extension of the \mssm\ with anomaly mediation as the
only source of \sy-breaking, and the tachyonic slepton problem solved by
a gauged  $\textrm{U(1)}$ symmetry. The extra gauge symmetry is broken
at high energies in a manner  preserving supersymmetry, while also
introducing both the see-saw mechanism for neutrino masses, and the
Higgs $\mu$-term. We call the model \samsb\ ({\it strictly\/} anomaly
mediated  \sy\ breaking. 
 
We present typical spectra for the model and compare them with those 
from so-called {\it minimal\/} anomaly mediated \sy\ breaking.  We find
a \sm-like Higgs of mass 125 GeV with a gravitino mass  of 140 TeV and
$\tan\beta=16$.  However, the muon anomalous magnetic
moment is 3$\sigma$ away from the experimental value. 

The model naturally produces a period of hybrid inflation, which can
exit to a false vacuum characterised by large Higgs vevs, reaching the
true ground state after  a period of thermal inflation. The
scalar spectral index is reduced to approximately 0.975, and the correct
abundance of neutralino dark matter can be produced by  decays of
thermally-produced gravitinos, provided the gravitino mass (and hence
the Higgs mass) is high. Naturally light cosmic strings are produced,
satisfying bounds from the Cosmic Microwave Background. The
complementary pulsar timing and cosmic ray bounds require that strings
decay primarily via loops into gravitational waves. Unless the loops are
extremely small, the next generation pulsar timing array will rule out
or detect the string-derived gravitational radiation background in this
model.}

\preprint{LTH 939, HIP-2012-10/TH\\\
}

\begin{document}

\section{Introduction}

The \sm\  Higgs-like particle of mass $125\GeV$ recently discovered at
the  LHC~\cite{:2012si,Chatrchyan:2012tx} strongly constrains future
model building, while recent negative results
from both the Tevatron and LHC in searches  for sparticles place
increasing pressure on models with low energy supersymmetry. Here we explore a specific
supersymmetric model in which the low energy spectrum is that of the
Minimal Supersymmetric Standard Model (\mssm), but the gauge symmetry is
augmented by an extra gauged  $\textrm{U(1)}$ symmetry,
$\textrm{U(1)}'$, spontaneously broken at high energies in a manner
which affects both physics at the supersymmetry breaking scale and
physics at high scales characterising inflation and cosmic strings. 

The broad features of the model are independent of the 
source of \sy\ breaking, but if we assume that this source is in fact 
anomaly mediation (\amsb)~\cite{Randall:1998uk}-\cite{Pomarol:1999ie}, 
then there arises an interesting interplay between 
the low energy physics (and in particular the Higgs $\mu$-term) and the 
high energy physics involving strings and inflation. Moreover the 
breaking of $\textrm{U(1)}'$ solves the tachyonic slepton problem characteristic of 
\amsb~\cite{Pomarol:1999ie,Jack:2000cd}.

We first presented this specific model in~\cite{Basboll:2011mh}, in a form where
we also introduced a Fayet-Iliopoulos (FI) term for the
$\textrm{U(1)}'$. Here we concentrate on the minimal formulation when
there is no such term\footnote{Aside from the fact that this makes the
model more appealing  by removing an independent mass scale, we also
thereby avoid confrontation with the conclusion of Komargodski and
Seiberg~\cite{Komargodski:2009pc}\ that a global theory with a FI term
cannot be consistently embedded in supergravity}. The model
implements a form of \amsb\ which we refer to as {\it strictly}
anomaly-mediated supersymmetry-breaking (\samsb), by which we mean  that
there are no other sources of supersymmetry breaking beside the  F-term
of the conformal compensator field. As a consequence the soft parameters
  have an elegant renormalisation group (\rg) invariant form. It
therefore differs from so-called  minimal \amsb (\mamsb), 
which posits an extra
source of supersymmetry-breaking, instead of  extra fields, in order to
solve the tachyonic  slepton problem. Our model is not quite a complete
\samsb\ implementation, in that  it requires an extension to determine
the soft parameter associated with the Higgs $\mu$-term.

We begin by describing the symmetries and field content of the model and
explaining in detail how the spontaneous breakdown of the
$\textrm{U(1)}'$ symmetry at a large scale $M$ not only solves the
\amsb\ tachyonic slepton problem, but also generates a Higgs $\mu$ term
and the see-saw mechanism for neutrino masses. This outcome is achieved
by the introduction of three  new chiral superfields; $S$, which is a
gauge singlet and a pair of  $\textrm{SU(3)}\otimes \textrm{SU(2)} \otimes \textrm{U(1)}_Y$ singlet
fields $\Phi, \Phib$ which are oppositely  charged under
$\textrm{U(1)}'$. We then exhibit characteristic sparticle spectra for
the model; the calculations involved to obtain these are essentially as
described in \references{Hodgson:2005en}{Hodgson:2007kq}, but allowing for a larger
gravitino mass. We also discuss the fine-tuning issue raised by this, 
and compare the results of our model with results from the most popular (but, we will 
argue, less elegant) version of \amsb, generally called \mamsb. 
We will see that \samsb\ generally keeps sleptons 
lighter than in \mamsb, which means that the contribution to the muon 
anomalous magnetic moment $a_\mu \equiv (g_\mu-2)/2$ is 
typically higher for a given Higgs mass.

The theory incorporates a natural mechanism for supersymmetric F-term
inflation, with the scalar component of $S$ as the inflaton. 
Previously, we concentrated on a region of parameter space such that
inflation ended with a transition to a state with only the 
$\textrm{U(1)}'$ broken. There is, however, an interesting alternative
that inflation ends with the development of vevs  for the Higgs
multiplets, $h_{1,2}$, breaking the electroweak symmetry.  A combination
of the Higgs fields $h_{1}\cdot h_{2}$ and the scalar components of the
singlet fields $\phi\phib$ is a flat direction,  lifted by soft
supersymmetry-breaking terms,  and the normal low energy electroweak
vacuum is achieved after a later period of  thermal inflation.  
Approximately 17 e-foldings of thermal inflation reduce the number  of
e-foldings of high scale inflation, and therefore reduce the spectral 
index of scalar Cosmic Microwave Background (CMB) fluctuations to about
0.975. This is within about $1\sigma$ of the WMAP7 value.

The reheat temperature after this period of thermal inflation is around
$10^9$ GeV, which means that  there is no gravitino problem: gravitinos
are very massive, more than 40 TeV, and so decay early enough not to be
in conflict with nucleosynthesis. Indeed, the gravitino problem can turn
into the gravitino solution for the typical \amsb\  feature of too low a
dark matter density generated at freeze-out: the lightest supersymmetric
particle (\lsp) is mostly wino and  has a relatively high annihilation
cross-section. In our model a critical density of {\lsp}s can be
generated by gravitino decays, if the gravitino (and hence the \lsp) is 
heavy enough. 

The model also has the possibility of baryogenesis via leptogenesis
following thermal inflation, with \cp\ violation supplied by the
neutrino sector. The field giving mass to right-handed neutrinos has an
inflation-scale ($10^{16}$ GeV) vacuum expectation value (vev), but if
the lightest right-handed neutrino is sufficiently light to be generated
at reheating after thermal inflation, a lepton asymmetry can be
generated by its out-of-equilibrium decay.

There is a broken U(1) symmetry in the model, and cosmic strings with a
$10^{16}$ GeV mass scale are formed, although not until the end of
thermal inflation.  There is a large Higgs condensate in the core of the
string which spreads the string out to a width of order the
supersymmetry-breaking scale, and reduces its mass per unit length by
well over an order of magnitude.  The strings satisfy CMB constraints on
the mass per unit length from combined WMAP7 and small-scale
observations. Their decays are constrained by pulsar timing observations
in the case of gravitational waves and the diffuse $\gamma$-ray
background in the case of particle production: the latter means that
less than about 0.1\% of the energy of the strings should end up as
particles, and the former puts constraints on the average size of the
loops at formation.

The model has the same field content as the $F_D$ hybrid inflation
model~\cite{Garbrecht:2006,Garbrecht:2006az}, 
but different charge assignments and
couplings. $F_D$ hybrid inflation also has a singlet which is a natural
inflaton candidate, but differs in other ways: for example, right-handed
neutrinos have electroweak-scale masses, and the gravitino problem is
countered by entropy generation.

The model also has the same field content as the B$-$L model of Refs.\
\cite{BuchmullerEtAl}, although as explained in section
\ref{sec:u1stuff}, the U(1)$'$ symmetry cannot be U(1)$_{\rm{B}-\rm{L}}$ in
\amsb.  It is also closely related to the model of
\reference{Dvali:1997uq}, in which the fields $\Phi$, $\Phib$ are
SU(2)$_R$ triplets. This also has a flat direction involving the Higgs,
although the authors did not pursue its consequences.

To summarise our results:  at $\tan\beta = 10$, \samsb\  can accommodate
a Higgs mass above 120 GeV for gravitino masses over 80 TeV, while
accounting for the  discrepancy in  $a_\mu$ between the Standard Model
(\sm) theory and experiment to  within $2\sigma$ would have  favoured 80
TeV or lower.  Larger values of $\tan\beta$ allow a more massive Higgs:
for $\tan\beta = 16$ we find a Higgs mass of 125 GeV  for a gravitino
mass of 140 TeV. 

\samsb\ also allows  
for an observationally consistent dark matter density, if the gravitino mass is 
over about 100 TeV, with the dark matter deriving from the decay of 
gravitinos produced from reheating after thermal inflation.  
The spectral index of scalar cosmological perturbations is within 1$\sigma$ of the WMAP7 value, 
and the observational bounds on cosmic strings can be satisfied if the strings decay into 
gravitational radiation.  The model has also
has a natural mechanism for baryogenesis via leptogenesis through the
decays of right-handed neutrinos.

\section{The \amsb{} soft terms}
\label{sec:soft}

We will assume that supersymmetry breaking arises via the renormalisation 
group invariant form characteristic of Anomaly Mediation, 
so that the soft parameters for the gaugino mass $M$, the $\phi^3$ interaction 
$h$ and the $\phi^*\phi$ and $\phi^2$ mass terms $m^2$ and $m_3^2$ in the \mssm\ 
take the generic \RG\ invariant form
\bea
M_i & = & m_{\frac{3}{2}} \beta_{g_i}/{g_i}\label{eq:AD1}\\
h_{U,D,E,N} & = & -m_{\frac{3}{2}}\beta_{Y_{U,D,E,N}}\label{eq:AD2}\\
(m^2)^i{}_j &=& \frac{1}{2}m_{\frac{3}{2}}^2\mu\frac{d}{d\mu}\gamma^i{}_j \label{eq:AD3}\\ 
m_3^2 & = & \kappa m_{\frac{3}{2}} \mu_h - m_{\frac{3}{2}} \beta_{\mu_h}.
\label{eq:AD4}
\eea
Here $\mu$ is the renormalisation scale, 
and $m_{\frac{3}{2}}$ is the gravitino mass; 
$\beta_{g_i}$ are the gauge $\beta$-functions and $\gamma$ is the 
chiral supermultiplet anomalous dimension matrix. 
$Y_{U,D,E,N}$ are the $3 \times 3$ Yukawa matrices, and $\mu_h$ is the superpotential Higgs $\mu$-term.  
We will see that in our low energy theory, \eqn{eq:AD3}\ is replaced, in fact, 
by
\ben
(m^2)^i{}_j = \frac{1}{2}m_{\frac{3}{2}}^2\mu\frac{d}{d\mu}\gamma^i{}_j
+kY_i\delta^i{}_j,\label{eq:AD3b} 
\een
where the $Y_i$ are charges corresponding to a $\textrm{U(1)}$ symmetry.
This $kY$ term corresponds in form to the contribution of an FI D-term,
and can be employed to  obviate the tachyonic scalar problem
characteristic of \amsb.   How such a term can be generated (with \amsb) was first
discussed in \reference{Pomarol:1999ie}, and  first applied to the MSSM
in  \reference{Jack:2000cd}.  The basic idea was pursed in  a number of
papers~\cite{ArkaniHamed:2000xj}-\cite{Jones:2006re}.  For example,
\reference{ArkaniHamed:2000xj}\  demonstrated explicitly the UV
insensitivity of the result, and  \reference{Murakami:2003pb}\
emphasised that the tachyonic problem  could be solved using a single
$\textrm{U(1)}$  rather than a linear combination of two, the approach
followed in \reference{Jack:2000cd}. An extension of the MSSM such that
the spontaneous breaking of a gauged $\textrm{U(1)}'$ with an  FI term
gave rise to the $kY$ term was written down in
\reference{Hodgson:2005en}. In~\cite{Basboll:2011mh}\ we developed an
improved version of this model,  retaining the possibility of a {\it
primordial\/} FI term for $\textrm{U(1)}'$;  here we will dispense with
the FI term, and emphasise that  we can nevertheless generate the $kY$
term naturally with $k$ of $O(m_\frac{3}{2}^2)$, by breaking a
$\textrm{U(1)}'$ symmetry at a large scale, {\it without introducing an
explicit FI term}. 

At first sight \eqn{eq:AD3b}\ resembles the formula for the scalar
masses employed in the so-called \mamsb\ model, where the $kY_i$ term
is replaced by a universal scalar mass contribution $m_0^2$. The
differences are as follows: 
\begin{itemize}

\item The \mamsb\ involves the introduction of
an additional  source of \sy\ breaking independent of the gravitino mass, 
while, as we shall see, \eqn{eq:AD3b}\
does not. 

\item The parameter $k$ in \eqn{eq:AD3b}\ turns out to be 
more constrained than $m_0^2$. This is associated with the fact that 
inevitably all the $Y_i$ cannot have
the same sign. 

\item The elegant \RG\ invariance of 
\eqn{eq:AD1}-\eqn{eq:AD4}\ is preserved by \eqn{eq:AD3b}.

\end{itemize}

It is these observations that prompts us to refer to our model as
\samsb.
Note that, of course,  we cannot ``promote"  the  \mamsb\ into the
\samsb\ by the addition of  additional heavier fields which cancel the
associated $\textrm{U(1)}'$ anomaly; with an  unbroken $\textrm{U(1)}'$,
any massive chiral multiplets  will obviously make no contribution to
this anomaly.

\eqn{eq:AD4} is the most general form for $m_3^2$ that is consistent
with  \rg\ invariance, as first remarked explicitly in
\reference{Hodgson:2005en};  the parameter $\kappa$ is an arbitrary
constant. For discussion of possible origins of $m_3^2$ from the
underlying superconformal  calculus formulation of supergravity see
\referencesb{Pomarol:1999ie}{Ibe:2004gh}{Jones:2006re}.   We will simply
assume that the model can be generalised  to produce such a term; the
procedure which has, in fact, been generally followed.  The presence of
$\kappa$ means that in sparticle spectrum calculations one is free to
calculate $m_3^2$ (and the value of the Higgs $\mu$-term, $\mu_h$) by
minimising the Higgs potential at the electroweak scale in the usual
way. (For $\kappa = 1$, which is the value suggested 
by a straightforward use of the conformal compensator field~\cite{Pomarol:1999ie}, 
one  might have hoped to use the minimisation
conditions to determine $\tan\beta$, but it turns  out this leads to a
very small value of $\tan\beta$ incompatible with gauge unification, because 
of the correspondingly large top Yukawa coupling~\cite{Kitano:2004zd}).  We
will see, however, that in our model the result for $\mu_h$ has
implications for other parameters in the {\it underlying\/} theory which
are constrained by cosmological considerations.

\section{The $\textrm{U(1)}'$ symmetry}
\label{sec:u1stuff}
The \mssm{} (including right-handed neutrinos) admits 
two independent generation-blind
anomaly-free $\textrm{U(1)}$ symmetries.
The possible charge assignments are shown in Table~\ref{anomfree}.
\begin{table}
\begin{center}
\begin{tabular}{|c|c c c c c c|} \hline
&$Q$ & $U$ & $D$
& $H_1$ & $H_2$ & $N$ \\ \hline
&& & & & & \\ 
$q$&$-\frac{1}{3}q_L$ & $-q_E-\frac{2}{3}q_L$ & $q_E+\frac{4}{3}q_L$
& $-q_E-q_L$ & $q_E+q_L$ & $-2q_L-q_E$ \\ 
&& & & & & \\ \hline
\end{tabular}
\caption{\label{anomfree}Anomaly free $\textrm{U(1)}$ symmetry for arbitrary
lepton doublet and singlet charges $q_L$ and $q_E$ respectively.}
\end{center}
\end{table} 
The \sm{} gauged $\textrm{U(1)}_Y$ is $q_L= - 1, q_E = 2$; this $\textrm{U(1)}$ is of
course anomaly free even
in the absence of $N$. $\textrm{U(1)}_{B-L}$ is
$q_E = -q_L = 1$; in the absence of $N$ this would have $\textrm{U(1)}^3$ and
$\textrm{U(1)}$-gravitational anomalies, but no mixed anomalies with the \sm{} gauge
group. 

Our model will have, in addition, a pair of \mssm{} singlet fields 
$\Phi, \Phib$ with $\textrm{U(1)}'$ charges
$q_{\Phi, \Phib} = \pm (4q_L+2q_E)$ and a gauge singlet $S$.
In order to solve the tachyon slepton problem we will need 
that, for our new gauge symmetry $\textrm{U(1)}'$, the charges $q_L,q_E$ have the 
same sign at low energies. As explained in \reference{{Hodgson:2007kq}}, however, 
it is in fact more appropriate to input parameters at high energies, when 
in fact although necessarily $q_E > 0$, the range of acceptable 
values of $q_L$ includes negative ones; not negative enough, however, to allow 
$\textrm{U(1)}'$ to be $\textrm{U(1)}_{B-L}$.

Thus \samsb\ has three  input parameters $m_{\frac{3}{2}}$, $kq_L$, $kq_E$, associated with 
the \sy\ breaking sector, while \mamsb\ only has two: $m_{\frac{3}{2}}$, $m_0$. However, 
it turns out that because the allowed $(q_L, q_E)$ region is so 
restricted, \samsb\ is the 
more predictive of the two. We will see this explicitly in section~\ref{sec:spectrum}.

\section{The superpotential and spontaneous $\textrm{U(1)}'$ breaking}

The complete superpotential for our model is:
\ben 
W = W_A + W_B
\label{eq:superpot}
\ee
where $W_A$ is the \mssm\ superpotential, omitting the Higgs 
$\mu$-term, and augmented by 
Yukawa couplings for the right-handed neutrinos, $N$:
\ben
W_A = H_2 Q Y_U U + H_1 Q Y_D D + 
H_1 L Y_E E + H_2 L Y_{N} N 
\ee
and
\ben
W_B = \lambda_1 \Phi\Phib S + \half\lambda_2 N N \Phi 
- \lambda_3 S H_1 H_2 - M^2 S,
\label{eq:wb}
\ee
where $M,\la_1,\la_3$ are real and positive 
and $\lambda_2$ is a symmetric $3\times 3$ matrix. The sign of the $\la_3$ term 
above is chosen because with our conventions, 
in the electroweak vacuum where 
\ben
H_1 = \left(\frac{v_1}{\sqrt{2}},0\right)^T\quad\hbox{and}\quad 
H_2 = \left(0, \frac{v_2}{\sqrt{2}}\right)^T
\een
we have $H_1 H_2 \to - \frac{1}{2}v_1 v_2$.

The $\textrm{U(1)}'$ symmetry forbids the renormalisable $B$ and $L$
violating superpotential interaction terms of the form $QLD$, $UDD$,
$LLE$, $H_1 H_2 N$, $N S^2$, $N^2 S$ and $N^3$, 
as well as the mass terms $N S$, $N^2$ and $L
H_2$ and the linear term $N$. Moreover $W_B$ contains the only 
cubic term involving $\Phi,\Phib$ that is allowed. 
Our superpotential \eqn{eq:superpot}\
is completely natural, in the sense that it is invariant under
a global $R$-symmetry, with superfield charges 
\ben
S=2, L=E=N=U=D=Q=1, H_1= H_2 = \Phi = \Phib = 0,
\label{rsym}
\ee
which forbids the remaining gauge invariant 
renormalisable terms ($S^2$, $S^3$, $\Phi\Phib$ and $H_1 H_2$). 
This $R$-symmetry also forbids the quartic superpotential
terms $QQQL$ and $UUDE$, which are allowed by the $\textrm{U(1)}'$ symmetry,
and give rise to dimension 5 operators capable of
causing proton decay~\cite{Weinberg:1981wj}-\cite{Sakai:1981pk}. 
It is easy to see, 
in fact, that the charges in \eqn{rsym}\ 
disallow B-violating operators in the superpotential 
of arbitrary dimension. 
Of course this $R$-symmetry is broken by the soft \sy\ breaking.
 
\section{The Higgs potential}
\label{s:HigPot}
In this section we discuss the spontaneous breaking of the
$\textrm{U(1)}'$ symmetry  and its consequences. We shall assume $M$ is
much larger than the  scale of \sy-breaking. (Such a large tadpole term
has been disfavoured in the past; moreover  it has been argued that it
would generally be expected to lead to a large  vev $\vev s$,  but as we
shall see this does not happen in our model.) It is then clear from the
form of the superpotential $W_B$  as given in \eqn{eq:wb} that for an
extremum that is \sic\ (when we neglect \sy-breaking)  we will require
non-zero  vevs for $\phi, \phib$ and/or $h_{1,2}$ (in order to obtain
$F_S = 0$).  The existence of competing vacua of this nature was noted
in by Dvali et al in  \reference{Dvali:1997uq}; their model differs from
ours in choice of gauge group (they  have $\textrm{SU(3)} \otimes
\textrm{SU(2)}_L \otimes \textrm{SU(2)}_R \otimes \textrm{U(1)}_{B-L}$)
and  \sy-breaking mechanism.

Let us consider these two possibilities in turn.

\subsection{The $\phi, \phib, s$ extremum}
\label{sec:phiphib}

Retaining for the moment only the scalar fields $\phi, \phib, s$ 
(the scalar component of their upper case counterpart 
superfields) we write the scalar potential:
\bea
V &=& \lambda_1^2 (|\phi s|^2 + |\phib s|^2) + |\lambda_1\phi\phib-M^2|^2
+\frak{1}{2}q_{\Phi}^2 g'^2\left(|\phi|^2 - |\phib|^2\right)^2\nn
&+&m_{\phi}^2 |\phi|^2 + m_{\phib}^2 |\phib|^2 +m_s^2|s|^2 
+\rho M^2 m_{\frac{3}{2}} (s + s^*)\nn
&+& h_{\lambda_1}\phi\phib s + c.c..
\label{eq:treepot}
\eea
Here, as well as soft terms dictated by \eqns{eq:AD2}{eq:AD3}, we also 
introduce a soft breaking  term linear in $s$. (In fact, according to
\reference{Jack:2001ew}, for a nonvanishing  {\it RG invariant\/}  
form of $\rho$  we would require a {\it quadratic\/} term in $s$ in the
superpotential, which in fact we do not have.  We nevertheless consider
the possible impact of a $\rho$ term, but will presently assume it is 
small, even if non-zero).

The potential depends on two explicit mass parameters, 
the gravitino mass $m_{\frac{3}{2}}$ and $M$.
Let us establish its minimum. Writing 
$\vev\phi = v_{\phi}/\sqrt{2}$,
$\vev\phib = v_{\phib}/\sqrt{2}$ and 
$\vev s = v_s/\sqrt{2}$, we find 
\bea
v_{\phi}\left[m_{\phi}^2+\half\la_1^2v_s^2
+\half g^2 q_{\Phi}^2(v_{\phi}^2-v_{\phib}^2)\right]
+v_{\phib}\left[\la_1\left(\half\lambda_1 v_{\phi} v_{\phib}-M^2\right)
+\frac{h_{\lambda_1}}{\sqrt{2}}v_s\right]&=&0
\label{eq:vphi}\\
v_{\phib}\left[m_{\phib}^2+\half\la_1^2v_s^2
-\half g^2 q_{\Phi}^2(v_{\phi}^2-v_{\phib}^2)\right]
+v_{\phi}\left[\la_1\left(\half\lambda_1 v_{\phi} v_{\phib}-M^2\right)
+\frac{h_{\lambda_1}}{\sqrt{2}}v_s\right]&=&0\label{eq:vphib}\\
v_s\left[m_s^2+\half\la_1^2(v_{\phi}^2+v_{\phib}^2)\right]
+\frac{h_{\lambda_1}}{\sqrt{2}}v_{\phi}v_{\phib}+\sqrt{2}\rho M^2 m_{\frac{3}{2}} &=&0.\label{eq:vs}
\eea

It follows easily from \eqns{eq:vphi}{eq:vphib} that 
\bea
\la_1\left(\half\lambda_1 v_{\phi} v_{\phib}-M^2\right)
&=& -\frac{v_{\phi}v_{\phib}}{v_{\phi}^2+v_{\phib}^2}
\left[m_{\phi}^2+m_{\phib}^2+\la_1^2 v_s^2\right] 
-\frac{h_{\lambda_1}}{\sqrt{2}}v_s\label{eq:Mterm}\\
\half {g'}^2 q_{\Phi}^2(v_{\phi}^2-v_{\phib}^2)
&=& \frac{v_{\phib}^2m_{\phib}^2-v_{\phi}^2m_{\phi}^2+ 
(v_{\phib}^2-v_{\phi}^2)\half\la_1^2v_s^2}{v_{\phi}^2+v_{\phib}^2}.\label{eq:gterm}
\eea
We now assume that $M \gg m_{\frac{3}{2}}$.
It is immediately clear from \eqns{eq:Mterm}{eq:gterm} that 
\ben
v_{\phi}^2 \simeq v_{\phib}^2 \simeq \frac{2}{\la_1}M^2
\label{eq:vsize}
\een
 and 
then from \eqn{eq:vs} that $v_{s}$ is $O(m_{\frac{3}{2}})$. 
We thus obtain from \eqn{eq:gterm} that 
\ben
v_{\phi}^2-v_{\phib}^2 = \frac{m_{\phib}^2-m_{\phi}^2}{{g'}^2q_{\Phi}^2} 
+ O(m_{\frac{3}{2}}^4/M^2)
\label{eq:vdiff}
\een
and from \eqn{eq:vs} that
\ben
v_s = -\frac{h_{\lambda_1}}{\sqrt{2}\la_1^2}-\frac{m_{\frac{3}{2}}\rho}{\sqrt{2}\la_1}
+ O(m_{\frac{3}{2}}^2/M). 
\label{eq:vsans}
\een
Now the $h_{\lambda_1}$ term is determined in accordance with 
\eqn{eq:AD2}:
\ben
h_{\lambda_1} = - m_{\frac{3}{2}}\frac{\la_1}{16\pi^2}\left(
3\la_1^2 +\frac{1}{2}\Tr \lambda_2^2 +2\lambda_3^2 - 4q_\phi^2{g'}^2 \right),
\een
denoting the $\textrm{U(1)}'$ charge by $g'$. 

If we assume that $|q_{\Phi}g'| \gg |\lambda_{1,2,3}|$ 
then we find 
\ben
v_s \simeq 
-\frac{\sqrt{2} \la_1 m_{\frac{3}{2}}}{16\pi^2}\left(\frac{2q_\phi^2{g'}^2}{\la_1^2}\right)
-\frac{m_{\frac{3}{2}}\rho}{\sqrt{2}\la_1}.
\label{eq:vsansb}
\een
For simplicity we shall assume that $|\rho| \ll q_\phi^2{g'}^2/(16\pi^2)$, 
so that 
the $\rho$ contributions
to \eqn{eq:vsans} and \eqn{eq:vsansb} are negligible.

Substituting back from \eqns{eq:vsize}{eq:vsans} into \eqn{eq:treepot}, we obtain to leading 
order 
\ben
V_{\phi} 
= \frac{1}{\la_1}M^2\left(m_{\phi}^2 +m_{\phib}^2 -\frac{h_{\la_1}^2}{2\la_1^2}\right)
\label{eq:phipot}.
\een
Presently we shall compare this result with the analagous one associated with the 
$h_{1,2},s$ extremum.

Supposing, however, that the $\phi,\phib,s$ extremum is indeed the relevant one, we obtain 
the Higgs $\mu$-term 
\ben
\mu_h = 
\frac{\la_1 \la_3 m_{\frac{3}{2}}}{16\pi^2}\left(\frac{2q_\phi^2{g'}^2}{\la_1^2}\right).
\label{eq:muans}
\een

One might think that since $v_s$ is naturally determined above to be
associated with the susy breaking scale (rather than the $\textrm{U(1)}'$
breaking scale) it would be necessary to minimise the whole Higgs
potential (including $\vev{h_{1,2}}$) in order to determine it. But if
we retain, for example, the $m_s^2$ term in \eqn{eq:vs}, the resulting
correction to \eqn{eq:vsans} is easily seen to be
$O(m_{\frac{3}{2}}^4/M^2)$. Similarly, the Higgs vevs responsible for
electroweak symmetry breaking do not affect \eqns{eq:vsans}{eq:muans} to
an appreciable extent. 

In \reference{Basboll:2011mh}, we naively estimated $\mu_h \sim \la_1 \la_3 m_{\frac{3}{2}}$, 
concluding that $\mu_h$ would be at most $O(\GeV)$
rather than $O(100\GeV)$. The improved formula \eqn{eq:muans}\ changes this conclusion.

If we neglect terms of $O(m_{\frac{3}{2}})$, it is easy to see from 
\eqns{eq:Mterm}{eq:gterm} that the breaking of $\textrm{U(1)}'$ preserves
supersymmetry (since in this limit the two equations correspond to
vanishing of the $S$ F-term and the $\textrm{U(1)}'$ D-term respectively);
thus the $\textrm{U(1)}'$ gauge boson, its gaugino (with one combination of
$\psi_{\phi,\phib}$) and the Higgs boson form a massive supermultiplet
with mass $m \sim g'\sqrt{v_{\phi}^2 + v_{\phib}^2}$, while the
remaining combination of $\phi$ and $\phib$ and the other combination
of $\psi_{\phi,\phib}$ form a massive chiral supermultiplet, with mass 
$m \sim \lambda_1\sqrt{v_{\phi}^2 + v_{\phib}^2}$.

For large $M$, all trace of the $\textrm{U(1)}'$ in the effective
low energy Lagrangian disappears, except for contributions to the masses
of the matter fields, arising from the $\textrm{U(1)}'$ D-term, which are
naturally of the same order as the \amsb{} ones. Evidently $S$ also
gets a large \sic\ mass, as does the $N$ triplet, thus naturally
implementing the see-saw mechanism. 
The generation of an appropriate $\mu$-term
via the vev of a singlet is reminiscent of the \nmssm{} (for a review
of and references for the \nmssm{} see Ref.~\cite{Ellwanger:2009dp}). We
stress, however, that our  model differs in a crucial way from the
\nmssm, in that the low energy spectrum is  precisely that of the \mssm.

It is easy to show by substituting \eqn{eq:vdiff} back into the
potential, \eqn{eq:treepot} that the contribution to the slepton masses
arising from the $\textrm{U(1)}'$ term which resolves the tachyonic slepton
problem is given by \ben \delta m^2_{l,e} \sim
\frac{q_{L,E}}{2q_{\Phi}}(m_{\phib}^2 - m_{\phi}^2) \label{eq:tachs}
\een with corresponding contributions for the other scalar MSSM fields 
proportional to their $\textrm{U(1)}'$ charges. Now \ben m_{\phib}^2 -
m_{\phi}^2 = 
\frac{1}{2}m_{\frac{3}{2}}^2\mu\frac{d}{d\mu}\left(\ga_{\phib}-\ga_{\phi}\right)
= -\frac{1}{2}\frac{m_{\frac{3}{2}}^2}{16\pi^2}\Tr
\lambda_2\beta_{\lambda_2} \een where (at one loop) \ben
16\pi^2\beta_{\lambda_2} = \lambda_2 \left[ \la_1^2 +2\lambda_2^2 +
\frac{1}{2}\Tr \lambda_2^2 +2 Y_N^{\dagger}Y_N -
(2q_\phi^2+4q_N^2){g'}^2\right] \label{eq:beta2}\\ \een and we have for
simplicity taken $\la_2$ to be diagonal. 

Let us consider what sort of values of $\delta m^2_{l,e}$ we require. 
In this context it is interesting to compare Fig.~1 of \reference{Hodgson:2005en} 
with Fig.~1 of \reference{Hodgson:2007kq}. In both references, $(L,e)$ correspond 
to our $(\delta m^2_l, \delta m^2_e)$ respectively. 
In the former case 
the scalar masses are calculated at low energies, whereas in the latter 
they are calculated at gauge unification and then run down to the electroweak scale. 
This is why the allowed $(L,e)$ regions are different in the two cases. 
Since we are assuming $M$ is large, it is clear that the latter are 
more relevant to our situation. From Fig.~1 of \reference{Hodgson:2007kq} we see that 
suitable values would be 
\ben
\delta m^2_l \simeq 0, \qquad 
0.16 \left(\frac{m_{\frac{3}{2}}}{40}\right)^2 \lesssim \delta m^2_e \lesssim 0.35 \left(\frac{m_{\frac{3}{2}}}{40}\right)^2.
\een 
Notice that $\delta m^2_e$ must necessarily be positive. 

So, if we assume that the one-loop $\beta_{\lambda_2}$ is dominated 
by its gauge contribution, consistent with our previous assumption that 
$|q_{\Phi}g'| \gg |\lambda_{1,2,3}|$, 
we obtain 
\ben
\delta m^2_{l,e} \simeq
\frac{q_{L,E}(q_{\Phi}^2+2q_N^2)}{2q_{\Phi}}\frac{{g'}^2
m_{\frac{3}{2}}^2\Tr \la_2^2}{(16\pi^2)^2}.
\label{eq:tach2}
\een
Now $q_{\Phi} = 4q_L + 2q_E$, so 
we see that it is easy to obtain the correct sign for $\delta m^2_e$.

For $q_L = 0$, we find 
\ben
\delta m^2_{e} \simeq
3q_E^2\frac{{g'}^2m_{\frac{3}{2}}^2\Tr \la_2^2}{2(16\pi^2)^2}
\label{eq:tach3}\een
or 
\ben
1.6 \lesssim q_E^2 {g'}^2\Tr \la_2^2 \lesssim 3.6.
\een
Of course with $q_L = 0$, we have $\delta m^2_{l}=0$; 
but as describe earlier, it was shown in \reference{Hodgson:2007kq}\ that 
acceptable slepton masses nevertheless result when we run down to low energies. 
Clearly there are similar contributions to the masses of the other matter 
fields similar to \eqn{eq:tach2}, thus for example
\ben
\delta m^2_{h_1,h_2} \simeq
\frac{q_{H_1,H_2}(q_{\Phi}^2+2q_N^2)}{2q_{\Phi}}\frac{{g'}^2
m_{\frac{3}{2}}^2\Tr \la_2^2}{(16\pi^2)^2}.
\label{eq:tach4}\een
In the notation of \reference{Hodgson:2007kq}, \eqn{eq:tach2}, for example, 
is simply replaced
by $\delta m^2_{l} = Lk'$ and $\delta m^2_{e} = ek'$ with $(L,e)$ replacing 
$q_{L,E}$, and all results presented for $k' = 1(\TeV)^2$.

We emphasise once again the contrast between our model and conventional 
versions of the \nmssm, which does not, in basic form, contain an extra 
gauged $\textrm{U(1)}$, but where a vev (of the scale of \sy\ breaking)
for the gauge singlet $s$ generates a Higgs $\mu$-term in much the same
way, as is done here. However, while in the \nmssm\ case the $s$ fields
are very much part of the Higgs spectrum, here, in spite of the
comparatively small $s$-vev, the $s$-quanta obtain large \sic\ masses
and are decoupled from the low energy physics, which becomes simply that
of the \mssm. Another nice feature is the natural emergence of the
see-saw  mechanism via the spontaneous breaking of the $\textrm{U(1)}'$.
Evidently  it will be feasible to associate  the $\textrm{U(1)}'$
breaking scale given by \eqn{eq:vsize}\ with the scale of gauge
unification. 

Although, as indicated above, we will be regarding $M$ as source of
significant  physics, it is worth briefly considering the limit $M \to
\infty$. In that limit, the theory becomes  simply the \mssm\ (including 
the Higgs $\mu_h$-term) with the
soft breaking terms given in \eqn{eq:AD1}-\eqn{eq:AD3} including the
additional  $kY$ term, which resolves the tachyon problem. The explicit
form of the terms proportional  to the gravitino mass in these equations
is easily derived using the conformal compensator field as described in
\reference{Pomarol:1999ie}.  Of course, although the resulting $kY$ term
in \eqn{eq:AD3}\ has the {\it form\/} of an FI term, in the effective
theory  (for $M \to \infty$) $U(1)'$ is not gauged and so we do not fall
foul of the strictures of \reference{Komargodski:2009pc}.  The conformal
compensator field does not provide us with a straightforward  derivation
of \eqn{eq:AD4}; as described  earlier, we will, like most previous
authors, rely on the electroweak minimisation process to determine the
Higgs $B$-term. 

\subsection{The $h_{1,2}, s$ extremum}.

We now consider the scalar potential
\bea
V &=& \lambda_3^2 (|h_1 s|^2 + |h_2 s|^2) + |\lambda_3 h_1 h_2-M^2|^2
+\frak{1}{2}g'^2 q_{H_1}^2\left(|h_1|^2 - |h_2|^2\right)^2\nn
&+& \frak{1}{8}g_1^2 (h_1^{\dagger}h_1
 -h_2^{\dagger}h_2)^2 + \frak{1}{8}g_2^2\sum_{a}(h_1^{\dagger}\sigma^a h_1
+h_2^{\dagger}\sigma^a h_2)^2\nn
&+&m_{h_1}^2 |h_1|^2 + m_{h_2}^2 |h_2|^2 +m_s^2|s|^2 
+\rho M^2 m_{\frac{3}{2}} (s + s^*)\nn
&+& h_{\lambda_3}h_1 h_2 s + c.c..
\label{eq:treepoth}
\eea
In \eqn{eq:treepoth}\ we have written the $U(1)_Y$ gauge coupling 
as $g_1$, although its normalisation corresponds to the usual \sm\ convention, not 
that appropriate for $\textrm{SU(5)}$ unification. This is to avoid confusion with the 
$\textrm{U(1)}'$ coupling, $g'$. 

We see that the potential is very similar to \eqn{eq:treepot}, the main difference 
being the presence of $\textrm{SU(2)}$ and $U(1)_Y$ D-terms. To leading order in $M$, only the 
$\textrm{SU(2)}$ D-term depends on the relative direction in $\textrm{SU(2)}$-space of the two doublets; 
it follows that we can choose without loss of generality to set 
$h_1 = (v_1/\sqrt{2},0)$ and $h_2 = (0,v_2/\sqrt{2})$, as in electroweak breaking, in 
order to obtain zero for the $\textrm{SU(2)}$ D-term for $v_1 = v_2$. 
Minimisation of the potential then proceeds in a similar way to the previous section 
(with the replacement $\la_1 \to \la_3$) leading to 
\ben
V_h = \frac{M^2}{\la_3}
\left(m_{h_1}^2+m_{h_2}^2-\frac{h_{\la_3}^2}{2\la_3^2}\right)
\label{eq:hpot}
\een
at the extremum. Here 
\bea
h_{\lambda_3} = - m_{\frac{3}{2}}\frac{\la_3}{16\pi^2}(
&\Tr& Y_E Y_E^{\dagger} + 
3\Tr Y_D Y_D^{\dagger}+ 3\Tr Y_U Y_U^{\dagger} 
+\lambda_1^2 + 4\Tr \lambda_3^2\nn
&-& 3g_2^2 - g_1^2 - 4q_{H_1}^2{g'}^2).
\eea
Let us compare the result for $V_h$ with that obtained for $V_{\phi}$, 
in the previous section, \eqn{eq:phipot}. If we assume that the $g'$ terms 
dominate throughout we obtain simply 
\ben
V_{\phi} = -\frac{M^2}{\la_1}\left(\frac{m_{\frac{3}{2}}{g'}^2}{16\pi^2}\right)^2
\left[4Q q_{\Phi}^2 + 8q_{\Phi}^4\right]
\label{eq:phivac}
\een 
and 
\ben
V_{h} = -\frac{M^2}{\la_3}\left(\frac{m_{\frac{3}{2}}{g'}^2}{16\pi^2}\right)^2
\left[4Q q_{H_1}^2 + 8q_{H_1}^4\right],
\label{eq:hvac}
\een 
where we have written the one loop $g'$ $\beta$-function as 
\ben
\beta_{g'} = Q \frac{{g'}^3}{16\pi^2}
\een
and 
\bea
Q &=& n_G (\frak{40}{3}q_L^2+8 q_E^2+16q_E q_L)+36q_L^2+40q_E q_L+12q_E^2\nn
&=& 76q_L^2+36q_E^2+88q_E q_L\quad (\hbox{for $n_G = 3$}).
\eea
The coefficient $Q$ is in general large, and larger than both
$q^2_\Phi$ and $q^2_{H_1}$, so the condition for the $\phi,\phib,s$
extremum to have a lower energy than the $h_1, h_2, s$ one may be written 
\ben
\label{e:EWCon}
\la_1\left(\frac{q_{H_1}}{q_{\Phi}}\right)^2 \lesssim \la_3.
\een
Alternatively, for the specific choice $q_L = 0$, which we will see in the 
next section leads to an acceptable electro-weak vacuum, we find 
that the same condition becomes
\ben
\label{e:EWConb}
\frac{19}{88}\la_1 \lesssim \la_3.
\een

\section{The sparticle spectrum}
\label{sec:spectrum}

In this section we calculate sparticle spectra for the \samsb\ model,
and compare the results with typical \mamsb\ spectra.  We shall be
interested  in seeking regions of parameter space with a ``high" Higgs
mass - that is, close to about 125 GeV as suggested  by recent LHC data
\cite{:2012si,Chatrchyan:2012tx} - and a  supersymmetric contribution to
the muon anomalous magnetic moment $\delta a_\mu$ compatible with
the experimental deviation from the  Standard  Model prediction, $\delta
a_{\mu}^{\textrm{exp}} = 29.5(8.8)\times 10^{-10}$ \cite{Miller:2007kk}.
 We will also wish to remain  consistent with the negative results of
recent LHC supersymmetry searches, see for example 
\references{Chatrchyan:2011zy}{Aad:2011ib}.

We use the methodology of \reference{Hodgson:2007kq}, which, as
explained in Section~\ref{sec:soft}, can also be applied to \mamsb\  by
replacing the characteristic $(L,e)$ FI-type terms of  \samsb\  by a
universal mass term $m_0^2$. 

We begin by choosing input values for $m_{\frac{3}{2}}$, $\tan\beta$, 
$L$, $e$ and
$\hbox{sign}\,\mu_h$ at the gauge unification scale $M_X$.
Then we calculate the appropriate
dimensionless coupling input values at the scale $M_Z$ by an iterative
procedure involving the sparticle spectrum, and the loop corrections to
$\alpha_{1\cdots 3}$, $m_t$, $m_b$ and $m_{\tau}$, as described in
Ref.~\cite{bpmz}. We define gauge unification by the meeting point of
$\alpha_1$ and $\alpha_2$; this scale, of around $10^{16}\GeV$, we
assume to be equal or close to the scale of $\textrm{U(1)}'$ breaking. For the
top quark pole mass we use $m_t = 172.9\GeV$. All calculations are done 
in the approximation that we retain only third generation Yukawa 
couplings, $\lambda_{t,b,\tau}$; thus the squarks and sleptons of the 
second generation are degenerate with the corresponding ones of the first 
generation. 

We then determine a given sparticle pole mass by running the
dimensionless couplings up to a certain scale chosen (by iteration) to
be equal to the pole mass itself, and then implementing full one-loop
corrections from \reference{bpmz}, and two-loop corrections to the top
quark mass~\cite{Bednyakov:2002sf}. We use two-loop anomalous dimensions
and $\beta$-functions throughout.

\subsection{Mass spectra in \samsb}

We display some examples of spectra in 
Tables~\ref{table:spectrumA}-\ref{table:spectrumD}. 
In each Table, the columns are
for different gravitino masses, all with $L=0$ with $e$ increasing with
increasing gravitino mass so as to remain within the allowed $(L,e)$
region; obviously $e$ scales like $m_{\frac{3}{2}}^2$ from
\eqn{eq:tach2}. (As already indicated, we input $(L,e)$ at $M_X$, so the
allowed $(L,e)$ region corresponds to that in
\reference{Hodgson:2007kq}\ rather than that in \reference{Hodgson:2005en}
). In Tables~\ref{table:spectrumA},\ref{table:spectrumC} 
the $(L,e)$ values are in the centre of the allowed
$(L,e)$ region (at least for smaller values of $m_{\frac{3}{2}}$), whereas 
in Tables~\ref{table:spectrumB},\ref{table:spectrumD} 
$e$ is smaller so  that lighter sleptons result.
We see that $\mu_h/m_{\frac{3}{2}}$ varies little 
with $m_{\frac{3}{2}}$; for example in Table~\ref{table:spectrumA} 
changing from $0.014$ at $m_{\frac{3}{2}}=40\TeV$ to 
$0.012$ at $m_{\frac{3}{2}}= 140 \TeV$. We thus find from 
\eqn{eq:muans}\ that 
\ben
\la_1 \la_3 \frac{2q_\phi^2{g'}^2}{\la_1^2} \simeq 2.2
\label{e:MuCon}
\een
in order for the electro-weak vacuum to exist. 
We shall return to this formula when 
we have discussed the cosmological constraints. 

In Tables~\ref{table:spectrumA}, \ref{table:spectrumB} we have 
$\tan\beta = 10$, 
whereas in Table~\ref{table:spectrumC},\ref{table:spectrumD} we have 
$\tan\beta = 16$.  Increasing $\tan\beta$ generally 
leads to a slight increase 
in the light Higgs mass $m_h$, and in the 
Table~\ref{table:spectrumD} case a much larger {\it decrease\/} in the 
heavy Higgs masses; this decrease is a signal of the fact that 
(for given $m_{\frac{3}{2}}$, $L$, $e$) there is an upper limit 
on $\tan\beta$; above that limit, the electroweak vacuum 
fails. 

Increasing the scale of \sy\ breaking (by increasing $m_{\frac{3}{2}}$) 
will, generally speaking, allow us to remain compatible with the more
stringent limits on BSM physics emerging from LHC searches and
$B$-decay. Recent LHC publications on supersymmetry searches (see for
example \references{Chatrchyan:2011zy}{Aad:2011ib}) tend to focus on 
sparticle spectra which are not compatible with \amsb; but it seems
clear that for  $m_{\frac{3}{2}} \gtrsim 60\TeV$ or so, our model is not
(yet)  ruled out. One search  result that explicitly targets anomaly
mediation is that of \reference{:2012jk}; this  sets a lower limit on
the wino mass of $92\GeV$, which in \samsb\ would correspond to 
$m_{\frac{3}{2}}\simeq 28\TeV$.

Increasing $m_{\frac{3}{2}}$ so as to reduce squark/gluino production
will,  however,   reduce the \sic\ contribution to the muon anomalous
magnetic moment $a_{\mu}$, and hence the opportunity to account for the
existing discrepancy between theory and experiment. But it is a feature
of \amsb, and in particular \samsb, that the sleptons are comparatively
light compared to the gluino and squarks. Therefore it turns out to be
possible to combine heavier coloured states  with sleptons and
electro-weak gauginos still light enough to contribute appreciably to
$a_{\mu}$.  We demonstrate this by including in the tables the result 
for the \sic\ contribution to $a_{\mu}$. For  $m_{\frac{3}{2}} =
60\TeV$,  the result is manifestly compatible with the afore-mentioned 
discrepancy.\footnote{ We use the one-loop formulae of
\reference{Moroi:1995yh}; for a review and  more references see
\reference{Stockinger:2006zn}.}

Notice that increasing $m_{\frac{3}{2}}$ so as to increase $m_h$ to
bring it closer  to the recent announcement of
evidence~\cite{:2012si,Chatrchyan:2012tx} for a \sm-like Higgs in the
region of $125\GeV$  can be done, but at the cost of reducing $\delta
a_{\mu}$; see the last column in Tables~\ref{table:spectrumC}, 
\ref{table:spectrumD}.   It also increases the degree  of fine-tuning,
as we shall discuss presently.

We can also increase  $\delta a_{\mu}$ by choosing $(L,e)$ closer to one
of the boundaries  of the allowed region corresponding to either the
charged slepton doublets or singlets   becoming too light; but the
effect of doing this is limited in that the gaugino masses are  not
sensitive to $(L,e)$. The bottom line is that with $\tan\beta = 10$, to
account for the whole of  $\delta a_{\mu}^{\textrm{exp}}$ we need a
light higgs mass of around  $115-120\GeV$. Increasing $\tan\beta$ also
leads to larger $\delta a_{\mu}$, but also a {\it smaller\/}  charged
Higgs mass, and a potentially over-large contribution to the branching
ratio $B \to X_s \gamma$. This effect is particularly noticeable in 
Table~\ref{table:spectrumC}, where the heavy Higgs masses actually 
{\it decrease\/} as $m_{\frac{3}{2}}$ is increased. We will return to this issue in
Section~\ref{sec:ft}. 

As in most versions of \amsb, the \lsp\ is mostly neutral wino, 
with the charged wino a few hundred MeV heavier.

\begin{table}\footnotesize
\begin{center}
\begin{tabular}{| c | c | c | c | c | c | c |} \hline
 $m_{\frac{3}{2}}$ & 40\TeV & 60\TeV & 80\TeV & 100\TeV & 120\TeV & 140\TeV
 \\ \hline
$ (L,e)$ & $(0,0.25)$ & $(0,0.5625)$ & $(0,1)$ & $(0,1.5625)$ & $(0,2.25)$ 
& $(0,3.0625)$ \\ \hline
$ {\tilde g}$ & 900 & 1297 & 1684 &  2062 & 2434 & 2802\\ \hline
$ \ttil_1$ & 757 & 1054 & 1346 & 1633 & 1915 & 2120\\ \hline
$ \ttil_2$ & 507 & 723 & 925 &  1115 & 1298 & 1473\\ \hline
$ \util_L $ & 819 & 1181 & 1531 & 1875 & 2211 & 2542 \\ \hline
$ \util_R $ & 766 & 1093 & 1408 & 1714 & 2012 & 2304 \\ \hline
$ \btil_1$ & 714 & 1023 & 1322 & 1614 & 1900 & 2181 \\ \hline
$\btil_2$ & 946 & 1376 & 1798 & 2213 & 2624 & 3031\\ \hline
$ \dtil_L$ & 822 & 1183 & 1533 & 1876 &  2212 & 2544\\ \hline
$\dtil_R$ & 955 & 1390 & 1816 & 2236 & 2651 & 3062 \\ \hline
$\tautil_1$ & 199 & 309 & 419 & 532 & 645 & 758\\ \hline
$\tautil_2$ & 266& 388& 512 & 635 & 759 & 882 \\ \hline
$\etil_L$ & 212 & 321 & 433  & 546 & 661 & 776  \\ \hline
$ \etil_R$ & 261 & 387 & 512 & 637 & 762 & 887\\ \hline
$\nutil_e $      & 249 & 378 & 506 & 632 & 758 & 883\\ \hline
$\nutil_{\tau} $ & 247 & 375 & 502 & 627 & 752 & 876\\ \hline
$\chi_1  $ & 131 & 198 & 265 & 331 & 396 & 461\\ \hline
$\chi_2  $ & 362 & 548 & 734 & 920 & 1107 & 1294\\ \hline
$\chi_3  $ & 588 & 841 & 1084 & 1319 & 1549 & 1773\\ \hline
$\chi_4  $ & 599 & 850 & 1091 & 1325 & 1552 & 1778\\ \hline
$\chi^{\pm}_1$ & 131& 199 & 265 & 331 & 396 & 461\\ \hline
$\chi^{\pm}_2$ & 597 & 848 & 1089 & 1324& 1552 & 1777\\ \hline
$h  $ & 115 & 118 & 120 & 122 & 123 & 124\\ \hline
$H,A$ & 366 & 492 & 595 & 680 & 749 & 802\\ \hline
$H^{\pm}$ & 374 & 499 & 601 & 685 & 753 & 806\\ \hline
$\chi^{\pm}_1-\chi_1$ (MeV) & {236} & {218} & {214} & {210} & {204} & 194
\\ \hline
$\mu_h$  & 571 & 812 & 1041 & 1259 & 1470 & 1675\\ \hline
$\delta a_{\mu}$  & $62 \times 10^{-10}$ & $26 \times 10^{-10}$ &
$13 \times 10^{-10}$ & $7.5 \times 10^{-10}$ & $4.6 \times 10^{-10}$
& $3.0 \times 10^{-10}$\\ \hline
\end{tabular}
\caption{\label{table:spectrumA}
\samsb\ mass spectra (in $\GeV$), and $\delta a_{\mu}$
for $m_t = 172.9\GeV$ and $\tan\beta = 10$}
\end{center}
\end{table}

\begin{table}\footnotesize
\begin{center}
\begin{tabular}{| c | c | c | c | c | c | c |} \hline
 $m_{\frac{3}{2}}$ & 40\TeV & 60\TeV & 80\TeV & 100\TeV & 120\TeV & 140\TeV
 \\ \hline
$ (L,e)$ & $(0,0.16)$ & $(0,0.36)$ & $(0,0.64)$ & $(0,1)$ & $(0,1.44)$ 
& (0,1.96)\\ \hline
$ {\tilde g}$ & 900 & 1297 & 1684 &  2063 & 2435 & 2802\\ \hline
$ \ttil_1$ & 770 & 1071 & 1369 & 1662 & 1951 & 2237\\ \hline
$ \ttil_2$ & 548 & 792 & 1023 &  1245 & 1460 & 1668\\ \hline
$ \util_L $ & 825 & 1191 & 1545 & 1892 & 2233 & 2568\\ \hline
$ \util_R $ & 795 & 1141 & 1474 & 1798 & 2116 & 2428\\ \hline
$ \btil_1$ & 723 & 1037 & 1342 & 1640 & 1933 & 2891\\ \hline
$\btil_2$ & 909 & 1320& 1721 & 2116 & 2506 & 2237\\ \hline
$ \dtil_L$ & 829 & 1194 & 1547 & 1894 &  2234 &  2922\\ \hline
$\dtil_R$ & 919 & 1334 & 1740 & 2140 & 2532  & 2569\\ \hline
$\tautil_1$ & 119 & 194 & 270 & 346 & 424 & 502\\ \hline
$\tautil_2$ & 198& 281& 366 & 452 & 537 & 623\\ \hline
$\etil_L$ & 145  & 219 & 295  & 373 & 452  & 532\\ \hline
$ \etil_R$ & 187 & 275 & 363 & 451 & 539  & 627\\ \hline
$\nutil_e $      & 170 & 263 & 354 & 444 & 533  & 622\\ \hline
$\nutil_{\tau} $ & 167 & 259 &  349 & 437 & 525 & 612\\ \hline
$\chi_1  $ & 131 & 198 & 265 & 330 & 395 & 460\\ \hline
$\chi_2  $ & 363 & 549 & 736 & 922 & 1109 & 1296\\ \hline
$\chi_3  $ & 635 & 916 & 1186 & 1450 & 1709 & 1964\\ \hline
$\chi_4  $ & 645 & 922 & 1192 & 1455 & 1713 & 1968\\ \hline
$\chi^{\pm}_1$ & 131& 199 & 265 & 330 & 395 & 460\\ \hline
$\chi^{\pm}_2$ & 643 & 921 & 1190 & 1454 & 1712  & 1967\\ \hline
$h  $ & 115 & 118 & 120 & 122 & 123  & 124\\ \hline
$H,A$ & 499 & 710 & 907 & 1094 & 1274 & 1448\\ \hline
$H^{\pm}$ & 506 & 716 & 911 & 1098 & 1277 & 1451\\ \hline
$\chi^{\pm}_1-\chi_1$ (MeV) & {223} & {213} & {212} & {210} & {204} & {197} \\ \hline
$\mu_h$  & 618 & 886 & 1142 & 1390 & 1470  & 1867\\ \hline
$\delta a_{\mu}$  & $62 \times 10^{-10}$ & $27 \times 10^{-10}$ &
$15 \times 10^{-10}$ & $9.2 \times 10^{-10}$ & $6.2 \times 10^{-10}$
  & $4.5 \times 10^{-10}$ \\ \hline
\end{tabular}
\caption{\label{table:spectrumB}
\samsb\ mass spectra (in $\GeV$), and $\delta a_{\mu}$
for $m_t = 172.9\GeV$ and $\tan\beta = 10$}
\end{center}
\end{table}

\begin{table}\footnotesize
\begin{center}
\begin{tabular}{| c | c | c | c | c | c |c|} \hline
 $m_{\frac{3}{2}}$ & 40\TeV & 60\TeV & 80\TeV & 100\TeV & 120\TeV & 140\TeV
 \\ \hline
$ (L,e)$ & $(0,0.25)$ & $(0,0.5625)$ & $(0,1)$ & $(0,1.5625)$ 
& $(0,2.25)$ & $(0,3.0625)$ \\ \hline
$ {\tilde g}$ & 899 & 1297 & 1683 &  2062 & 2434 & 2801\\ \hline
$ \ttil_1$ & 750 & 1041 & 1328 & 1612 & 1892 & 2168\\ \hline
$ \ttil_2$ & 504 & 721 & 924 &  1116 & 1300 & 1745\\ \hline
$ \util_L $ & 819 & 1181 & 1532 & 1875 & 2211  & 2543\\ \hline
$ \util_R $ & 766 & 1094 & 1409 & 1714 & 2012 & 2305\\ \hline
$ \btil_1$ & 703 & 1007 & 1301 & 1590 & 1873  & 2153\\ \hline
$\btil_2$ & 929 & 1352 & 1768 & 2177 & 2582 & 2983\\ \hline
$ \dtil_L$ & 823 & 1183 & 1534 & 1876 &   2213 & 2544\\ \hline
$\dtil_R$ & 955 & 1391 & 1812 & 2236 & 2651  & 3062\\ \hline
$\tautil_1$ & 182 & 291 & 400 & 511 & 621  & 733\\ \hline
$\tautil_2$ & 271& 391 & 512 & 633 & 755 & 877\\ \hline
$\etil_L$ & 212 & 321 & 433  & 546 & 660  & 776\\ \hline
$ \etil_R$ & 262 & 387 & 512 & 638 & 762  & 887\\ \hline
$\nutil_e $      & 249 & 378 & 506 & 632 & 758  & 883\\ \hline
$\nutil_{\tau} $ & 244 & 372 & 497 & 621 & 752 & 867\\ \hline
$\chi_1  $ & 132 & 199 & 265 & 331 & 396 & 461\\ \hline
$\chi_2  $ & 362 & 548 & 734 & 920 & 1107 & 1294\\ \hline
$\chi_3  $ & 585 & 836 & 1077 & 1311 & 1539 & 1763\\ \hline
$\chi_4  $ & 594 & 843 & 1083 & 1316 & 1544 & 1767\\ \hline
$\chi^{\pm}_1$ & 132& 199 & 265 & 331 & 396 & 461\\ \hline
$\chi^{\pm}_2$ & 592 & 842 & 1082 & 1315& 1543  & 1766\\ \hline
$h  $ & 116 & 119 & 121 & 123 & 124  & 125\\ \hline
$H,A$ & 284 & 366 & 417 & 440 & 430& 374\\ \hline
$H^{\pm}$ & 285 & 375 & 425 & 447 & 438 & 384\\ \hline
$\chi^{\pm}_1-\chi_1$ (MeV) & {229} & {216} & {213} & {210} & {204} & {194}\\ \hline
$\mu_h$  & 566 & 806 & 1032 & 1249 & 1458  & 1662\\ \hline
$\delta a_{\mu}$  & $1 \times 10^{-8}$ & $41 \times 10^{-10}$ &
$21 \times 10^{-10}$ & $12 \times 10^{-10}$ & $7.5 \times 10^{-10}$
  & $4.9 \times 10^{-10}$ \\ \hline
\end{tabular}
\caption{\label{table:spectrumC}
\samsb\ mass spectra (in $\GeV$), and $\delta a_{\mu}$
for $m_t = 172.9\GeV$ and $\tan\beta = 16$}
\end{center}
\end{table}

\begin{table}\footnotesize
\begin{center}
\begin{tabular}{| c | c | c | c | c | c |c|} \hline
 $m_{\frac{3}{2}}$ & 40\TeV & 60\TeV & 80\TeV & 100\TeV & 120\TeV & 140\TeV
 \\ \hline
$ (L,e)$ & $(0,0.18)$ & $(0,0.405)$ & $(0,0.72)$ & $(0,1.125)$ & $(0,1.62)$ 
& (0,1.96)\\ \hline
$ {\tilde g}$ & 899 & 1297 & 1684 &  2062 & 2434 & 2801\\ \hline
$ \ttil_1$ & 761 & 1056 & 1348 & 1635 & 1918 & 2199\\ \hline
$ \ttil_2$ & 536 & 775  & 1001 &  1218 & 1426 & 1629\\ \hline
$ \util_L $ & 928 & 1191 & 1543 & 1889 & 2228  & 2563\\ \hline
$ \util_R $ & 824 & 1131 & 1460 & 1780 & 2094 & 2402\\ \hline
$ \btil_1$ & 710 & 1019 & 1348 & 1611 & 1898 & 2181\\ \hline
$\btil_2$ & 901 & 1309 & 1708 & 2101 & 2488 & 2872\\ \hline
$ \dtil_L$ & 828 & 1192 & 1545 & 1890 &   2230 & 2564\\ \hline
$\dtil_R$ & 928 & 1347 & 1757 & 2161 & 2559  & 2954\\ \hline
$\tautil_1$ & 111 & 196 & 280 & 364 & 448  & 532\\ \hline
$\tautil_2$ & 223& 314 & 405 & 498 & 590 & 683\\ \hline
$\etil_L$ & 163 & 245 & 331 & 418 & 508 & 595\\ \hline
$ \etil_R$ & 206 & 304 & 401 & 499 & 596  & 693\\ \hline
$\nutil_e$      & 190 &     293  & 393 & 492 & 591 & 689\\ \hline
$\nutil_{\tau}$ & 184  & 284 & 381 & 477  & 573 & 668\\ \hline
$\chi_1  $ & 132 & 199 & 265 & 330 & 396 & 460\\ \hline
$\chi_2  $ & 363 & 549 & 735 & 922 & 1108 & 1295\\ \hline
$\chi_3  $ & 621 & 892 & 1156 & 1412 & 1664 & 1910\\ \hline
$\chi_4  $ & 630 & 898 & 1161 & 1417 & 1668 & 1914\\ \hline
$\chi^{\pm}_1$ & 132& 199 & 265 & 331 & 396 & 460\\ \hline
$\chi^{\pm}_2$ & 628 & 898 & 1160 & 1416 & 1667  & 1913\\ \hline
$h  $ & 116 & 119 & 121 & 123 & 124  & 125\\ \hline
$H,A$ & 410 & 577 & 729 & 869 & 1001 & 1125\\ \hline
$H^{\pm}$ & 419 & 583 & 734 & 873 & 1005  & 1129\\ \hline
$\chi^{\pm}_1-\chi_1$ (MeV) & {218} & {212} & {211} & {209} & {204} & {195}\\ \hline
$\mu_h$  & 603 & 863 & 1111 & 1379 & 1584  & 1852\\ \hline
$\delta a_{\mu}$  & $101 \times 10^{-10}$ & $44 \times 10^{-10}$ &
$24 \times 10^{-10}$ & $14 \times 10^{-10}$ & $9.5 \times 10^{-10}$
  & $6.5 \times 10^{-10}$ \\ \hline
\end{tabular}
\caption{\label{table:spectrumD}
\samsb\ mass spectra (in $\GeV$), and $\delta a_{\mu}$
for $m_t = 172.9\GeV$ and $\tan\beta = 16$}
\end{center}
\end{table}

\subsection{Comparison with \mamsb}

It is interesting to compare the \samsb\ spectra presented 
in Tables~\ref{table:spectrumA}-\ref{table:spectrumD}\ 
with some \mamsb\ spectra. 

\begin{table}\footnotesize
\begin{center}
\begin{tabular}{| c | c | c | c | c |} \hline
 $m_0$ & 450 & 900 & 1800 & 2700
 \\ \hline
$ {\tilde g}$ & 1310 & 1342 & 1398 & 1438 \\ \hline
$ \ttil_1$ & 1156 & 1303 & 1783 & 2398 \\ \hline
$ \ttil_2$ & 940 & 1052 & 1384 & 1804 \\ \hline
$ \util_L $ & 1295 & 1499 & 2135 & 2912 \\ \hline
$ \util_R $ & 1285 & 1489 & 2126 & 2903\\ \hline
$ \btil_1$ & 1120 & 1278 & 1773& 2392 \\ \hline
$\btil_2$ & 1288 & 1489 & 2121 & 2892 \\ \hline
$ \dtil_L$ & 1287 & 1491 & 2128  & 2904\\ \hline
$\dtil_R$ & 1303 & 1506 & 2141  & 2917 \\ \hline
$\tautil_1$ & 355 & 851 & 1764 & 2664\\ \hline
$\tautil_2$ & 399& 870& 1774 & 2671\\ \hline
$\etil_L$ & 381 & 865 & 1778& 2680\\ \hline
$ \etil_R$ & 390 & 871 & 1784 & 2687\\ \hline
$\nutil_e $ & 372 & 861 & 1776 & 2679 \\ \hline
$\nutil_{\tau} $ & 367 & 856 & 1768 & 2668 \\ \hline
$\chi_1  $ & 199 & 200 & 201& 202\\ \hline
$\chi_2  $ & 550 & 555 & 558 & 559 \\ \hline
$\chi_3  $ & 1031 & 1027 & 1004 & 950 \\ \hline
$\chi_4  $ & 1037 & 1032 & 1009 & 956 \\ \hline
$\chi^{\pm}_1$ & 200& 201 & 201 & 202 \\ \hline
$\chi^{\pm}_2$ & 1036 & 1031 & 1009 & 955 \\ \hline
$h  $ & 118 & 119 & 120 & 122 \\ \hline
$H,A$ & 1076 & 1314 & 2006 & 2802\\ \hline
$H^{\pm}$ & 1079 & 1317 & 2008 & 2804 \\ \hline
$\chi^{\pm}_1-\chi_1$ (MeV) &  209 & 209  & 208 & 209 \\ \hline
$\mu_h$  & 1000 & 989 & 956 & 889 \\ \hline
$\delta a_{\mu}$  & $22 \times 10^{-10}$ & $6.1 \times 10^{-10}$ & 
$0.57 \times 10^{-10}$ & 0.10  $\times 10^{-10}$
 \\ \hline
\end{tabular}
\caption{\label{table:spectrumE}
\mamsb\ mass spectra (in $\GeV$), and $\delta a_{\mu}$ for $m_{\frac{3}{2}}=60\TeV$, $m_t = 172.9\GeV$ and $\tan\beta = 10$} 
\end{center}
\end{table}

In Table~\ref{table:spectrumE}\ we present results for 
$m_{\frac{3}{2}}=60\TeV$, for different values of $m_0$.  The second
column of this table corresponds to the Benchmark Point  \mamsb 1.3 of
\reference{AbdusSalam:2011fc}; our  results for the masses  agree
reasonably well with those presented there: for example, the gluino
masses  differ by 2\%, and the lightest third generation squarks by 1\%.
They are also not inconsistent with those of \reference{Arbey:2011ab},
who quote an  upper limit for $m_h$ of $120.4\GeV$; note that there the
parameter  scan is restricted to $m_0 < 2\TeV$. For a detailed
comparison of \mamsb\ results with  recent LHC data see
\reference{Allanach:2011qr}.  We see that by increasing $m_0$, we can
eventually make  all the squarks heavier than the the gluino; this is 
not possible in \samsb, because increasing $L$ and $e$ soon leads to 
loss of the electro-weak vacuum. We will discuss this fact in more
detail in  Section~\ref{sec:ft}. 

In Table~\ref{table:spectrumF}\ we present the corresponding results for
$m_{\frac{3}{2}}=140\TeV$.  Note the (comparitively) light 
sleptons in column 2 of this
Table; these occur because  for these values the $m_0^2$ contribution to
the slepton $({\rm masses})^2$  almost cancels the  (negative)
$m_{\frac{3}{2}}^2$ one. (We do not give results in Table~\ref{table:spectrumF}\ 
for $m_0 = 450\GeV$, because in that case there are still 
tachyonic  sleptons). This is analagous to being close to a boundary
in  the allowed $(L,e)$ space in the \samsb\ case, and, as there, does
not in itself result in a large  $\delta a_{\mu}$, because the wino
masses are unaffected. Moreover, away from the $(L,e)$ boundary  (in
\samsb) the slepton masses remain relatively small, whereas for fixed
$m_{\frac{3}{2}}$, increasing $m_0$ (in \mamsb) leads rapidly to larger slepton
masses.

\begin{table}\footnotesize
\begin{center}
\begin{tabular}{| c | c | c | c |} \hline
 $m_0$ & 900 & 1800 & 2700
 \\ \hline
$ {\tilde g}$ & 2824 & 2881 & 2939 \\ \hline
$ \ttil_1$ & 2382 & 2682 & 3114 \\ \hline
$ \ttil_2$ & 2038 & 2248 & 2548\\ \hline
$ \util_L $ & 2776 & 3162 & 3720 \\ \hline
$ \util_R $ & 2756 & 3143 & 3703\\ \hline
$ \btil_1$ & 2362 &  2668  & 3105 \\ \hline
$\btil_2$ & 2704 & 3085 & 3636 \\ \hline
$ \dtil_L$ & 2757 & 3144 & 3707\\ \hline
$\dtil_R$ & 2795 & 3179 & 3735 \\ \hline
$\tautil_1$ & 620 & 1652 & 2573\\ \hline
$\tautil_2$ & 710 & 1691 & 2605\\ \hline
$\etil_L$ & 707 & 1717 & 2634\\ \hline
$ \etil_R$ & 723 & 1707 & 2643\\ \hline
$\nutil_e $ & 703 & 1705 & 2632 \\ \hline
$\nutil_{\tau} $ & 670 & 1678 & 2560 \\ \hline
$\chi_1  $ & 461 & 464 & 465\\ \hline
$\chi_2  $ & 1297 & 1306 & 1311 \\ \hline
$\chi_3  $ & 2240 & 2211 & 2162 \\ \hline
$\chi_4  $ & 2243 & 2214 & 2164\\ \hline
$\chi^{\pm}_1$ & 461 & 464 & 465 \\ \hline
$\chi^{\pm}_2$ & 2242 & 2213 & 2164 \\ \hline
$h  $ & 125 & 126 & 126 \\ \hline
$H,A$ & 2186 & 2618 & 3214\\ \hline
$H^{\pm}$ & 2188 & 2620 & 3216 \\ \hline
$\chi^{\pm}_1-\chi_1$ (MeV) & 191  & 175 & 161 \\ \hline
$\mu_h$  & 2136 & 2095 & 2032 \\ \hline
$\delta a_{\mu}$  & $5.8 \times 10^{-10}$ &
$0.95 \times 10^{-10}$ & $0.22  \times 10^{-10}$
 \\ \hline
\end{tabular}
\caption{\label{table:spectrumF}
\mamsb\ mass spectra (in $\GeV$), and $\delta a_{\mu}$ for 
$m_{\frac{3}{2}}=140\TeV$,
$m_t = 172.9\GeV$ and $\tan\beta = 16 $}
\end{center}
\end{table}

It is interesting that in \mamsb, increasing $m_0$ (for fixed 
$m_{\frac{3}{2}}$) leads  to a slight decrease in $\mu_h$, and a
consequent slight  decrease in the  masses of the heavy neutralinos and
chargino. Note also that  the \sic\ contribution to $a_{\mu}$ is
compatible with  $\delta a_{\mu}^{\textrm{exp}}$ for  $m_0 = 450\GeV$, in 
Table~\ref{table:spectrumE}, but
decreases rapidly as $m_0$ increases.  If we increase $m_{\frac{3}{2}}$ to 
$140\TeV$ as in Table~\ref{table:spectrumF}, we are 
able to obtain $m_h = 125\GeV$, but, as in \samsb\,  at the price of a small 
contribution to $\delta a_{\mu}$.

\subsection{Fine tuning}
\label{sec:ft}

Noting that as $m_{\frac{3}{2}}$ is increased we find that $\mu_h$
increases, we should comment on the issue of the fine-tuning required to
produce the electro-weak scale. From the well-known tree level relation

\ben
\frac{m_{h_1}^2-m_{h_2}^2\tan^2\beta}{\tan^2\beta -1} -\mu_h^2 = \frak{1}{2}M_Z^2
\een
we see that unless $|m_{h_1}^2| \gg |m_{h_2}^2|$ then,
for typical values of $\tan\beta$, we have
\ben
\mu_h^2 \simeq -m_{h_2}^2 - \frak{1}{2}M_Z^2
\een
which since generically $|m_{h_2}| \gg M_Z$ represents a fine tuning, sometimes
called the ``little hierarchy'' problem.

One might have hoped, since $q_{H_2} = q_L + q_E > 0$, to reduce
$|m_{h_2}^2|$, and hence $\mu_h^2$, by increasing $q_L + q_E$; see
\eqn{eq:tach4}. But from Fig.~1 of \reference{Hodgson:2007kq}\ we see
$L+e$ is severely constrained by the requirement of a stable electroweak
vacuum; the failure of this is manifested by a tachyonic $m_A$.
The tree formula for $m_A$ is
\ben
m_A^2 = 2\mu_h^2 + m_{h_1}^2 + m_{h_2}^2 \simeq m_{h_1}^2 - m_{h_2}^2 -M_Z^2 
\een
and it is apparent from \eqn{eq:tach4}\ that the overall effect of increasing
$q_L + q_E$ actually decreases $m_A^2$.

For example, if we use $m_{\frac{3}{2}}= 80\TeV$ and $(L,e)$ =
$(0,1.2)$, then we find that $m_A$ is sharply reduced to $295\GeV$ while
$\mu_h$ changes only to $980\GeV$. A small further increase in $e$ takes
$m_A$ rapidly to zero. A similar outcome is the result of increasing
$\tan\beta$. For example, with $m_{\frac{3}{2}}= 80\TeV$ and $(L,e)$ =
$(0,1)$, as in the fourth column of Table~\ref{table:spectrumA}, $\mu_h$
decreases with increasing $\tan\beta$ but $m_A$ decreases more sharply.
For $\tan\beta = 19$, we find $\mu_h = 1025\GeV$, but $m_A = 207\GeV$,
and for $\tan\beta = 20$, $m_A^2 < 0$.

If we increase $m_{\frac{3}{2}}$ then the upper limit on $\tan\beta$ 
decreases; for example with $m_{\frac{3}{2}}= 120\TeV$ and $(L,e) =
(0,2.25)$ as in the  sixth column of Table~\ref{table:spectrumA}, we
find that the maximum value of $\tan\beta$ is $\tan\beta = 17$,  with
$m_h = 124\GeV$ and $m_A = 309\GeV$, and $\delta a _{\mu} = 7.9 \times
10^{-10}$. Note that $\delta a_{\mu}$ increases as  $\tan\beta$
increases; however, in Table~\ref{table:spectrumC}, 
the concomitant decrease in the Higgs masses (in
particular the  charged Higgs mass) leads to an increased \sic\
contribution to the branching ratio for  $B \to X_s\gamma$, and
potential conflict with experiment. See Figure~4 of 
\reference{Allanach:2009ne}. This problem is avoided in 
Table~\ref{table:spectrumD}; but with $m_{\frac{3}{2}}$ large enough 
to produce $m_h = 125\GeV$, there is no region in $(L,e)$ space 
permitting a $\tan\beta$ large enough to generate 
$\delta a_{\mu} \approx 20-30 \times 10^{-10}$.


Within the context of our model we see no clean way to avoid the fine-tuning 
problem. It is interesting to note that with the alternative GUT-compatible
assignment considered in Section~5 of \reference{Hodgson:2007kq}, $L +
e$ can be increased if desired (see Fig.~2 of that reference). However
in that case we have $q_{H_1} = -e-L$ and $q_{H_2} = -2e$, 
so increasing $L + e$ does not reduce $|m^2_2|$ or $m_A^2$.

\FIGURE{
\includegraphics[scale=0.5]{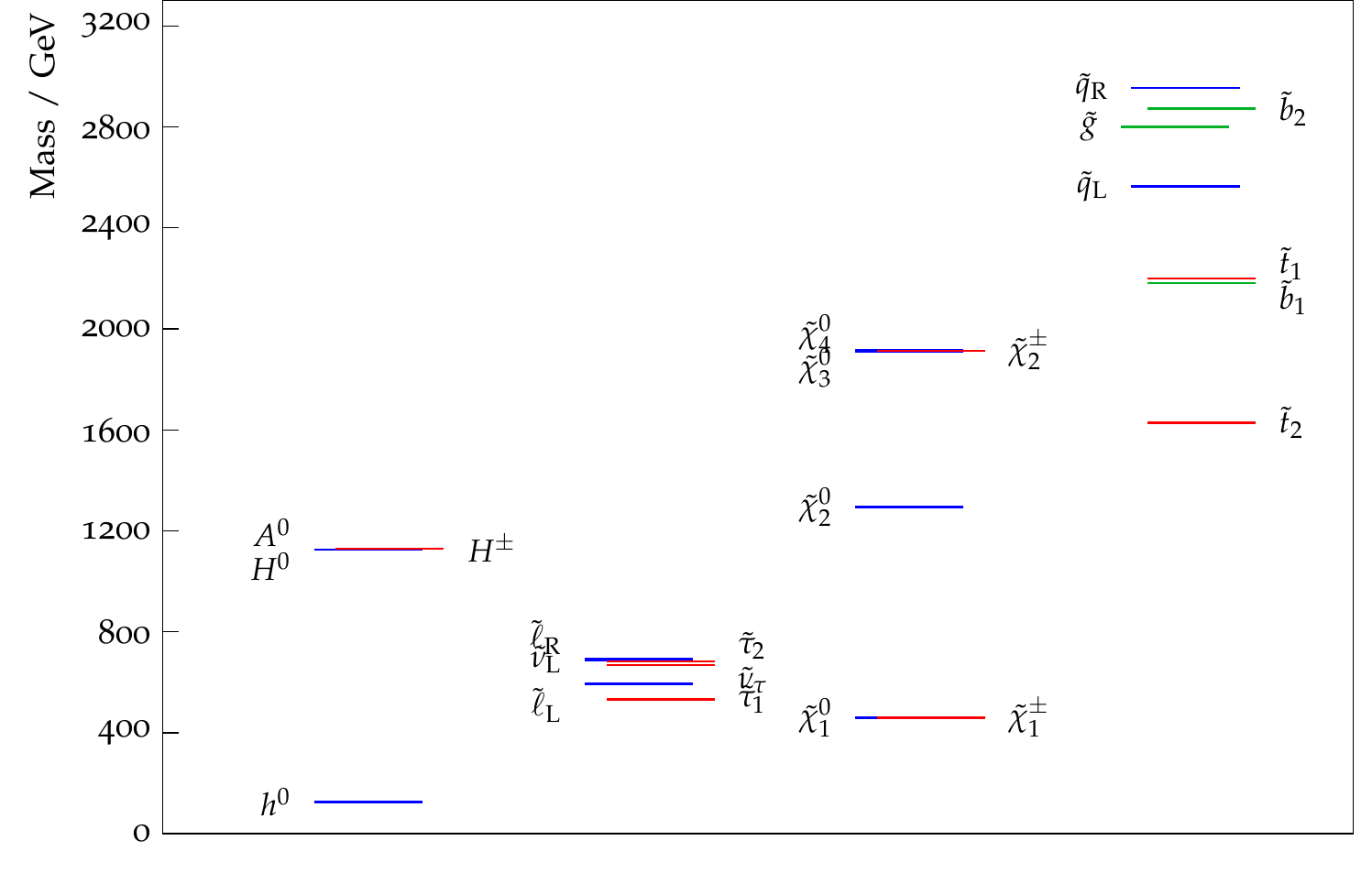}
\includegraphics[scale=0.5]{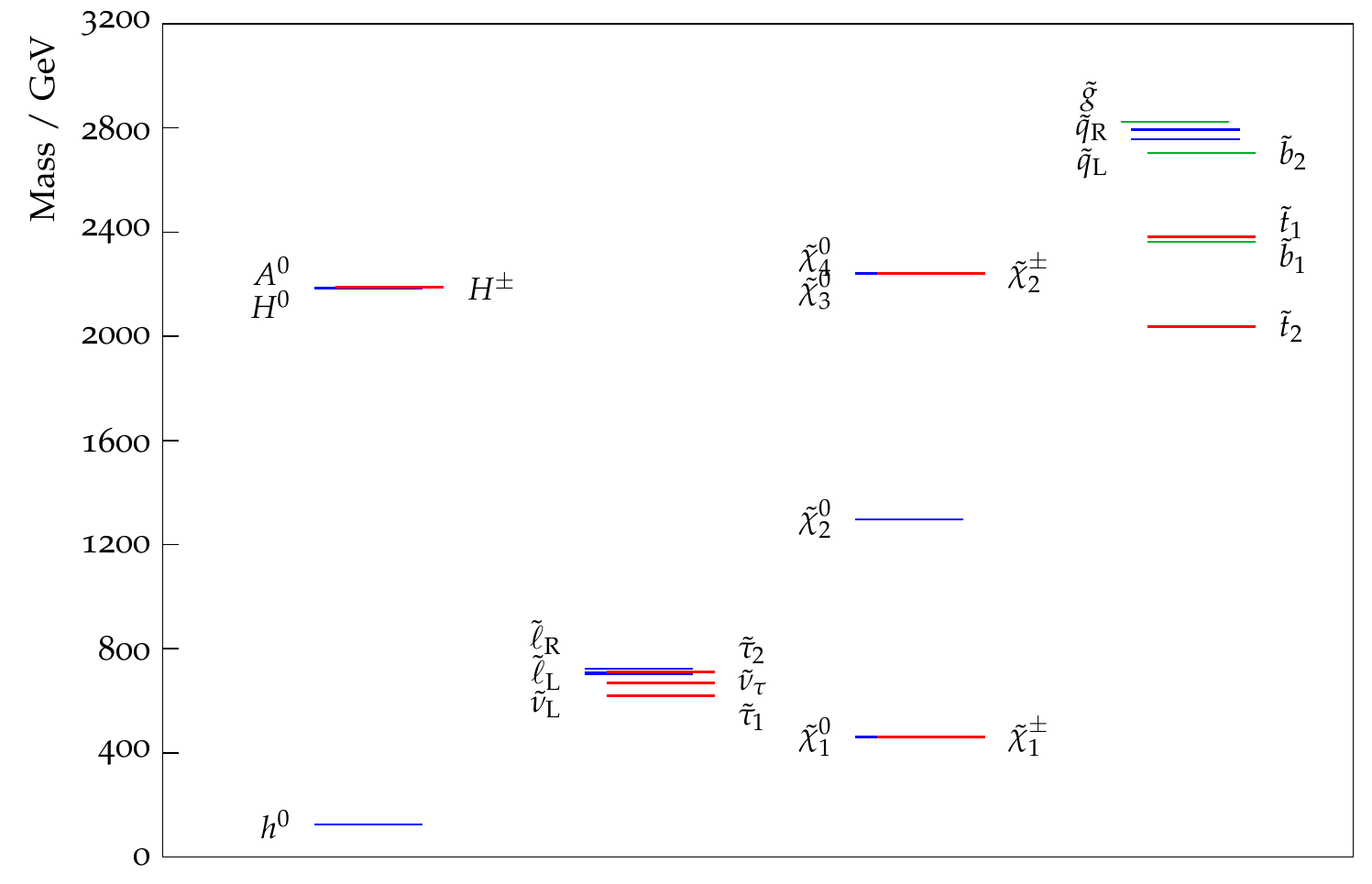}
\includegraphics[scale=0.5]{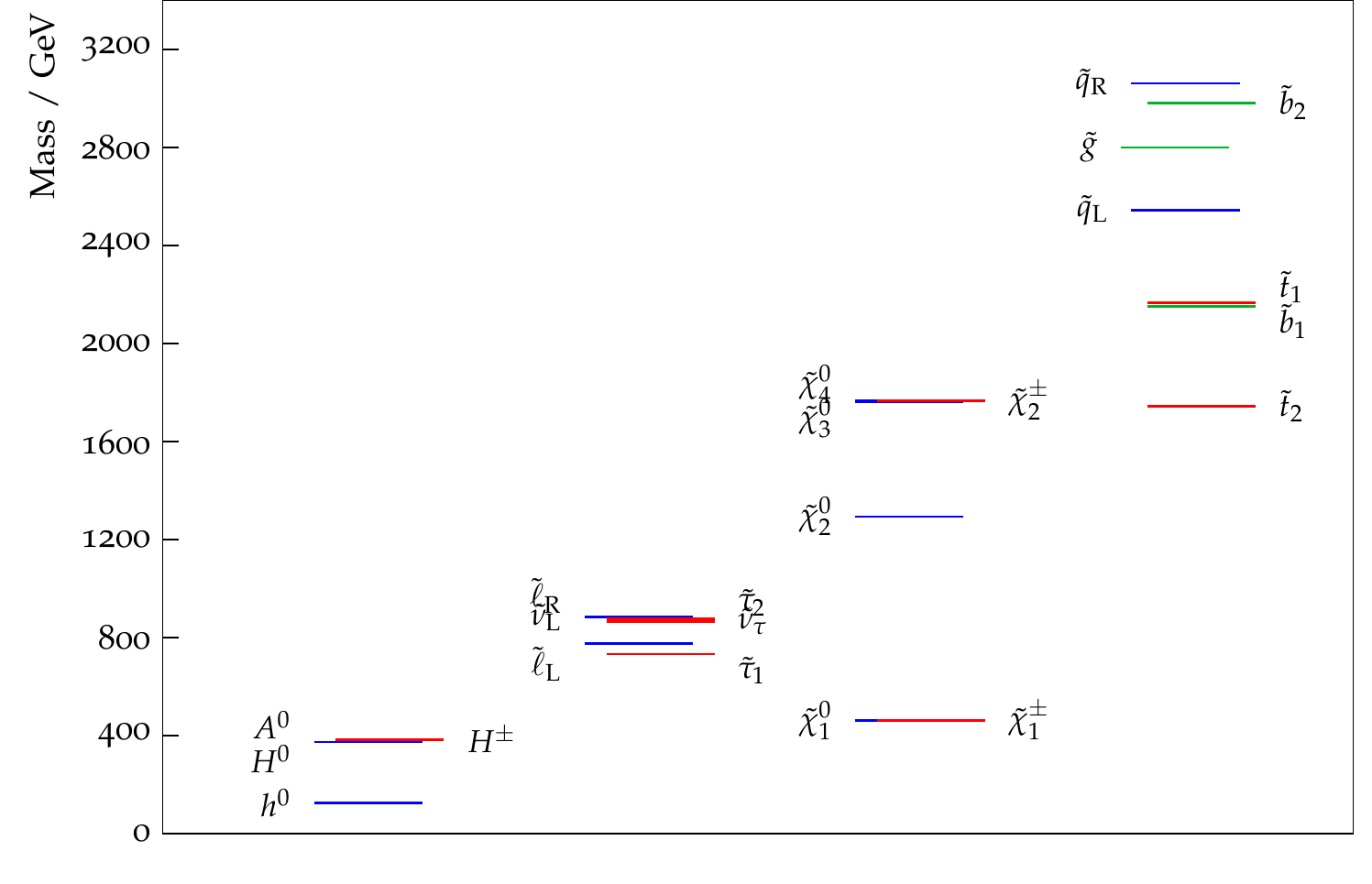}
\includegraphics[scale=0.5]{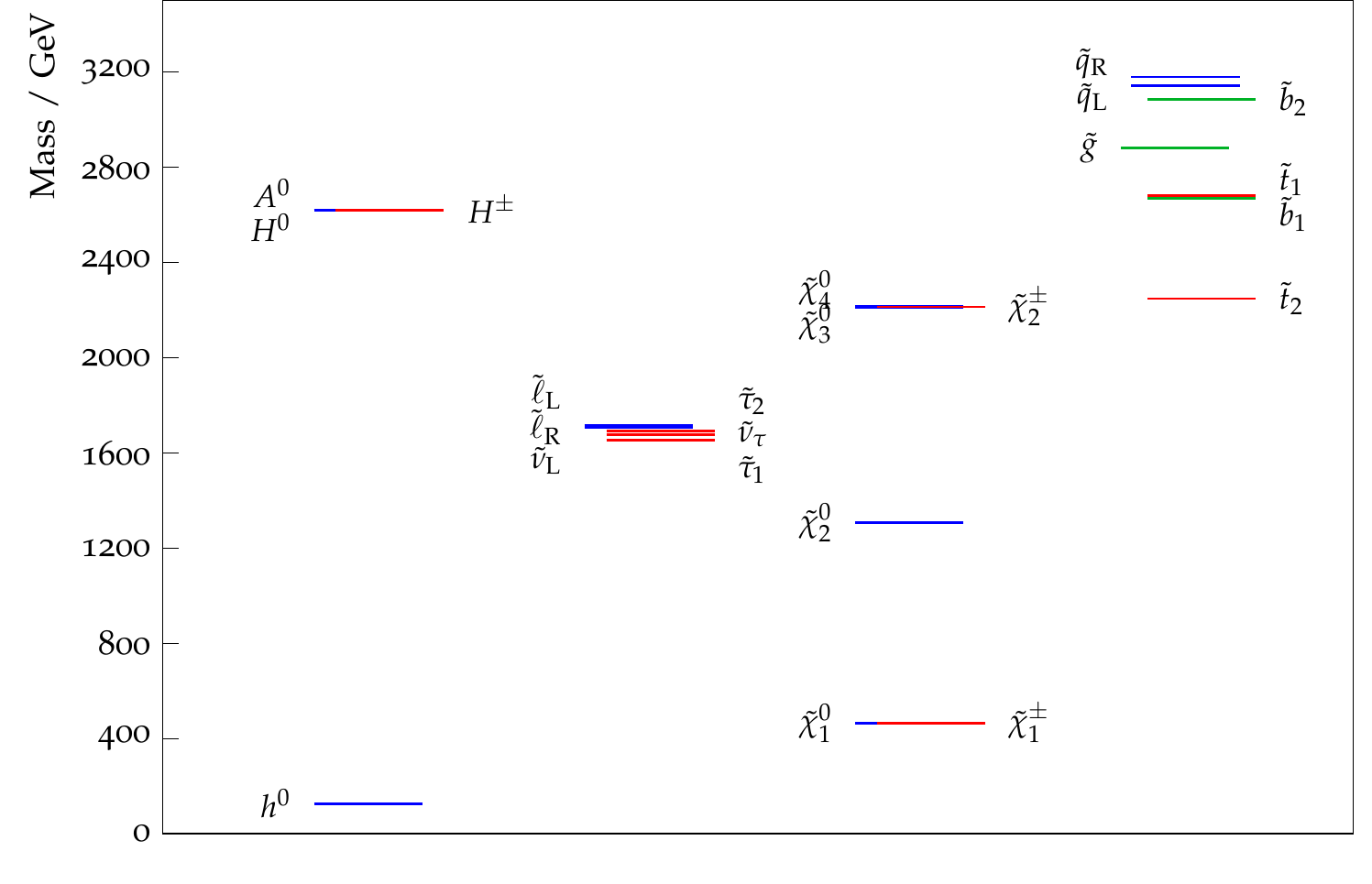}
\caption{
Comparison between \samsb\ (left) and \mamsb\ (right) mass spectra, 
drawn from column 7 of 
Tables \ref{table:spectrumC} and \ref{table:spectrumD} (left), and 
columns 2 and 3 of Table
\ref{table:spectrumF} (right). The gravitino masses are 140 TeV, and $\tan\beta = 16$ in all cases. 
The resulting Higgs masses are between 125 GeV  and 126 GeV.
Note how the increase in the magnitude of the sAMSB D-term contribution to the soft masses decreases the masses of the non-SM Higgs particles.
\label{f:spectra} 
}
}

\section{Cosmological history}

\label{effpot}

\subsection{F-term inflation}

As detailed in a previous paper \cite{Basboll:2011mh}, the theory
naturally produces F-term 
inflation~\cite{Copeland:1994vg}-\cite{Lyth:1998xn}, with the singlet
scalar $s$ as the inflaton. In this paper we are assuming that the
FI-term vanishes, which considerably simplifies the radiative
corrections driving the evolution of $s$ during inflation. 
We also assume that the quartic term in $s$ in the K\"ahler potential is 
negligible. 

The relevant terms in the tree potential are
\bea
V_{\rm tree} &=& |\lambda_1\phi\phib - \lambda_3 h_1 h_2 - M^2|^2 + 
\left[\lambda_1^2 (|\phi|^2 + |\phib|^2) 
+ \lambda^2_3 (|h_1 |^2 +|h_2 |^2)\right] | s |^2
\nn 
&+&\frac{1}{2}g'^2\left(q_{\Phi}(\phi^*\phi -\phib^*\phib)
+ q_{H_1} (h_1^{\dagger}h_1- h_2^{\dagger}h_2)\right)^2\nn
&+& \frak{1}{8}g_2^2 \sum_{a}(h_1^{\dagger}\sigma^a h_1 
+h_2^{\dagger}\sigma^a h_2 )^2 + \frak{1}{8}g_1^2 ( h_1^{\dagger}h_1
 - h_2^{\dagger}h_2)^2\nn
&+& V_\mathrm{soft},
\label{eq:treepotb}
\eea
where we have used $q_{\bar\Phi} = -q_\Phi$, $q_{H_2} = - q_{H_1}$ arising 
from the anomaly cancellation and gauge invariance conditions.
The \amsb{} soft terms 
$V_\mathrm{soft}$ are the sum of those appearing in \eqns{eq:treepot}{eq:treepoth}, and are all suppressed 
by at least one power of $m_{\frac{3}{2}}$, 
which we are assuming to be much less than $M$.
The most important soft term is the linear one, which we are assuming is absent or at least small (see the discussion following Eq.\ \ref{eq:treepot}).

 
At large $s$ and vanishing $\phi$, $\phib$, $h_1$ and $h_2$, and neglecting soft terms,  we have 
\ben
V_\mathrm{tree} = M^4 + \Delta V_1,
\ee
where $\Delta V_1$ represents the one-loop corrections, given as usual by 
\ben
\Delta V_1 = \frac{1}{64 \pi^2} 
{\rm Str} \left[(M^2(s))^2\ln(M^2(s)/\mu^2)\right].
\label{Vone}
\ee
Here 
\ben
{\rm Str} \equiv \sum_{\rm scalars} - 2 \sum_{\rm fermions} 
+ 3 \sum_{\rm vectors}.
\ee
In the absence of the FI term, $\Delta V$ is in fact 
dominated by the $\Phi$, $\Phib$ and $H_{1,2}$ subsystems, and 
the contribution to the one-loop scalar potential is \cite{Basboll:2011mh}
\bea
\Delta V_1 
&=&\frac{1}{32\pi^2}\left[(\lambda_1^2 s^2+\lambda_1 M^2)^2\ln 
\left(\frac{\la_1^2s^2+\lambda_1 M^2}{\mu^2}\right)
+ (\lambda_1^2 s^2-\lambda_1 M^2)^2\ln 
\left(\frac{\la_1^2s^2-\lambda_1 M^2}{\mu^2}\right) \right.\nn 
&+& 2(\lambda_3^2 s^2+\lambda_3 M^2)^2\ln 
\left(\frac{\la_3^2s^2+\lambda_3 M^2}{\mu^2}\right)
+ 2(\lambda_3^2 s^2-\lambda_3 M^2)^2\ln 
\left(\frac{\la_3^2s^2-\lambda_3 M^2}{\mu^2}\right)\nn
&-&\left. 2\lambda_1^4 s^4\ln\left(\frac{\la_1^2s^2}{\mu^2}\right)
 -4\lambda_3^4 s^4\ln\left(\frac{\la_3^2s^2}{\mu^2}\right)
\right].
\eea
For values of $s$ for which $\la_{1,3} s^2 \gg M^2$ 
it is easy to show that, after removing a finite local counterterm, this reduces to 
\ben 
V(s) \simeq M^4 \left[1+ \al\ln
\frac{2s^2}{s_c^2}\right], 
\label{e:VsApp}
\een 
where   
\ben 
\alpha = \frac{\lambda^2}{16\pi^2}, \quad \la = 
\sqrt{\lambda_1^2+ 2\lambda_3^2}, \quad  s_c^2 = M^2/\lambda.
\label{eq:newalpha}
\een
Note that neglecting the linear soft term $\rho M^2 m_\frac32 s + \textrm{c.c.}$ is equivalent to assuming
\ben
\label{e:RhoCon}
\rho \ll \frac{\la^3}{16\pi^2} \frac{s_c}{m_\frac32}.
\een
With the parameterisation (\ref{e:VsApp}), the scalar and tensor power spectra $\powspec_s$, $\powspec_t$ and the scalar spectral index $n_s$ generated $N$ e-foldings before the end of inflation are 
\bea 
\powspec_s(k) & \simeq & \frac{1}{24\pi^2} \frac{2N_k}{\al}
\left(\frac{M}{m_p}\right)^{4} = \frac{4N_k}{3}  \left(\frac{s_c}{m_p}\right)^{4}, \\ 
\powspec_t(k) &\simeq & 
\frac{1}{6\pi^2}\left(\frac{M}{m_p}\right)^{4} = \frac83 \left(\frac{s_c}{m_p}\right)^{4} , \\ 
n_s & \simeq & \left(1 -
\frac{1}{N_k}\right). 
\eea 
The WMAP7 best-fit values for $\powspec_s(k_0)$ and $n_s$ at $k = k_0 = 0.002\;h\textrm{Mpc}^{-1}$ 
in the standard $\La$CDM model are \cite{Komatsu:2010fb}
\ben\label{eq:WMAP7ps}
\powspec_s(k_0) = (2.43\pm0.11) \times 10^{-9}, \quad n_s = 0.963\pm0.012 (68\% \textrm{CL}),
\een
which correspond to 
\ben
\frac{s_c}{m_p} \simeq 2.9 \times 10^{-3} \left(\frac{27}{N_{k_0}}\right)^\frac14, \quad N_{k_0} = 27^{+13}_{-7}. 
\een
There is an approximately 2$\si$ discrepancy with the standard Hot Big Bang result
$N_{k_0} \simeq 55 + \ln(T_\mathrm{rh}/10^{15}\;\mathrm{GeV})$.  We will see later how this is ameliorated by $N_\th \simeq 17$ e-foldings of thermal inflation, reducing the discrepancy to approximately 1$\si$.

If $\la_3 > \la_1$, inflation ends at 
the critical value 
$
 s^2_{c1} = M^2/\la_1,
$
followed by transition to the $\textrm{U(1)}'$-broken phase described by 
\eqn{eq:vphi}-\eqn{eq:vs}. 
On the other hand, if $\lambda_3 < \lambda_1$, we find that $s_{c3}^2 =
M^2/\lambda_3$, and the
Higgses develop vevs of order the unification scale rather than $\phi, \phib$. 

At first sight this rules out this latter possibility and in \cite{Basboll:2011mh} we
did not explore it. However, we saw in Section
\ref{s:HigPot} that the condition for the correct (small Higgs vev)
electroweak vacuum to have the lowest energy density (\ref{e:EWCon}) is
slightly less restrictive than the condition for inflation to exit to
the $\phi$-$\phib$ direction, and that there is a range of parameters
\ben
 \la_1\left(\frac{q_{H_1}}{q_{\Phi}}\right)^2 \lesssim \la_3 < \la_1
\label{eq:la3range}
\een
for which the universe exits to the false high Higgs vev $h$-vacuum. It
then should evolve to the true ground state: in this section we will see
that this evolution leads to a very interesting cosmological history,
with some distinctive features. 

\subsection{Reheating} \label{sec:reheating}
If inflation exits to the $h$-vacuum the symmetry-breaking is 
\ben
\textrm{SU(2)}\times\textrm{U(1)}_Y\times \textrm{U(1)}' \quad \to \quad 
\textrm{U(1)}_\textrm{em}\times\textrm{U(1)}'',
\een
where the U(1)$''$ is generated by the linear combination of 
hypercharge and U(1)$'$ 
generators 
which leaves the Higgses invariant: 
\ben
Y'' = Y - \frac{Y'}{q_L + q_E}.
\een
Topologically, the symmetry-breaking is the same as in the Standard Model, and hence cosmic strings are not formed at this transition. 

Reheating after hybrid inflation \cite{PreHeaHybInf} is expected in our model to be very rapid, 
as the non-perturbative field interactions of the scalars with fermions \cite{Berges:2010zv} and with gauge fields \cite{DiazGil:2005qp} are very efficient 
at transferring energy out of the zero-momentum modes of the fields $s$,  $h_1$ and $h_2$. 
Higgs modes decay rapidly into $b$ quarks, leading to the universe regaining a relativistic equation of state in much less than a Hubble time.
Hence the universe thermalises at a temperature $T_\textrm{rh} \simeq M$.  

One notices that before thermal effects and soft terms are taken into
account, the minimum of the scalar potential is determined by the
requirement that both the F- and D-terms vanish. The vanishing of the
D-terms ensures that $|{\phi}| = |{\phib}|$, $|h_1| = |h_2|$ and
$h_1^\dag h_2 = 0$, while the vanishing of the F-term is assured by
$\lambda_1\phi\phib - \lambda_3 h_1 h_2 = M^2$.  The minimum can
therefore be parametrized by an SU(2) gauge transformation and angles
$\chi,\varphi$ defined by 
\bea
\vev{h_1}& \simeq& - i\si_2\vev{h_2}^* \simeq (\frac{M}{\sqrt\la_3}\cos\chi,0), 
\nn
\vev{\phi} &\simeq & \vev{{\phib}^*} 
\simeq \frac{M}{\sqrt{\la_1}}\sin\chi e^{i\varphi}.
\eea
The $\varphi$ angle can always be removed by a U(1)$''$ gauge
transformation, so the physical flat direction just maps out the
interval $0 \le \chi \le \pi/2$. At the special point $\chi = 0$ the
U(1)$''$ symmetry is restored, and at $\chi=\pi/2$ the
$\textrm{SU(2)}\otimes\textrm{U(1)}_Y$ is restored. Away from these special points only
U(1)$_\textrm{em}$ is unbroken.

With this parametrisation, it is straightforward to show that the leading O($M^2m_\frac32^2$) terms in the effective potential for $\chi$ are, after solving for $s$,
\ben
V(\chi) \simeq -\frac{M^2}{2}\frac{\left( \tilh_{\la_1}\sin^2\chi + \tilh_{\la_3}\cos^2\chi\right)^2}{\la_1\sin^2\chi +{\la_3}\cos^2\chi}
+ M^2 \left( \frac{\barm_\phi^2}{\la_1}\sin^2\chi + \frac{\barm_h^2}{\la_3}\cos^2\chi \right),
\een
where we have defined $\tilh_{\la_1} = \frac{h_{\la_1}}{\la_1}$, $\tilh_{\la_3}=\frac{h_{\la_3}}{\la_3}$, $\barm_\phi^2 = m_\phi^2 + m_{\bar\phi}^2$ and $\barm_h^2 = m_{h_1}^2 + m_{h_2}^2$.
A little more algebra demonstrates that
\ben
V''(0) \simeq \frac{2M^2}{\la_3}\left[-\frac{\tilh_{\la_3}^2}{2} \left(2 \frac{\tilh_{\la_1}}{\tilh_{\la_3}} - \frac{\la_1}{\la_3} - 1\right)  + \barm_\phi^2\frac{\la_3}{\la_1} - \barm_h^2\right],
\een
while the expansion around the true vacuum (the $\phi$-vacuum) 
at $\chi=\pi/2$ is easily obtained by the replacements $1 \leftrightarrow 3$ and $\barm^2_\phi \leftrightarrow \barm^2_h$.

In \samsb\ we have, under our assumption that the U(1)$'$ couplings
dominate the $\beta$-functions, 
\ben \tilh_{\la_1} \simeq \left(
\frac{m_\frac32{g'}^2}{16\pi^2}\right) 4q_\phi^2, \quad \tilh_{\la_3}
\simeq \left( \frac{m_\frac32{g'}^2}{16\pi^2}\right) 4q_{H_1}^2, \een
and 
\ben \barm_h^2 \simeq - \left(
\frac{m_\frac32{g'}^2}{16\pi^2}\right)^2 4Qq_{H_1}^2, \quad 
\barm_\phi^2 \simeq - \left( \frac{m_\frac32{g'}^2}{16\pi^2}\right)^2
4Qq_{\phi}^2. \een 
As pointed out in Section \ref{s:HigPot}, $Q$ is in
general much larger than both $q_{\phi}^2$ and $q_{H_1}^2$, so we see
that the $h$-vacuum is unstable only if 
\ben
\barm^2_\phi\frac{\la_3}{\la_1} - \barm_h^2 \lesssim 0, 
\een or 
\ben
\frac{\la_3}{\la_1} \gtrsim \frac{q_{H_1}^2}{q_\phi^2}. 
\een 
This
coincides with the condition (\ref{e:EWCon}) that the $h$-vacuum has
higher energy than the $\phi$-vacuum, and that the $\phi$-vacuum is
stable.

Note that we can define a canonically normalised U(1)$''$-charged complex scalar modulus 
field $X$, related to $\chi$ and $\varphi$  in the neighbourhood of the
$h$-vacuum by 
\ben
X \simeq \sqrt\frac{2M^2}{\la_1} \chi e^{i\varphi}
\een
and whose mass $m_X$ is given by
\ben
\label{e:Xmas}
m_X^2 \simeq \barm_\phi^2\frac{\la_3}{\la_1} - \barm_h^2.
\een

\subsection{High temperature ground state}

As we outlined in the previous section, reheating is expected to take
place in much less than a Hubble time $H \sim M^2/\mpl$, while the
relaxation rate to the true ground state, the $\phi$ vacuum, is from
\eqn{e:Xmas} $m_X$. Given that we expect $m_X \sim 1$ TeV and $M \sim
10^{14}$ GeV, reheating happens much faster than the relaxation, and the
universe is trapped in the U(1)$''$-symmetric vacuum with the large
Higgs vev.

The high temperature effective potential, or free energy density, can be written 
\ben
f(X,T) = -\frac{\pi^2}{90}g_\textrm{eff}(X,T)T^4,
\een
where $g_\textrm{eff}(X,T)$ is the effective number of relativistic
degrees of freedom at temperature $T$. 
At weak coupling, $g_\textrm{eff}(X,T)$ can be calculated in the
high-temperature expansion for all particles of mass $m_i \ll T$
\cite{Dolan:1973qd}, 
\ben
g_\textrm{eff}(X,T) \simeq g^0_\textrm{eff} - \frac{90}{\pi^2} \sum_i  {c_{1,i}}  \frac{m_i^2(X)}{T^2}, \quad 
\een
where $g^0_\textrm{eff}$ is the  effective number of degrees of freedom at $X=0$,  
and $c_1 = \frac{1}{24},\frac{1}{48}$ for bosons and fermions respectively. 
For particles with $m > T$, $g_\textrm{eff}$ is exponentially suppressed.

We can see that $X = 0$ is a local minimum for temperatures $m_{X} \lesssim T \lesssim M$, because away from that point 
the  U(1)$''$ gauge boson develops a mass $q_\phi g'|X|$, and so $g_\textrm{eff}$ decreases. 
For similar reasons the $\phi$-vacuum at $X_\phi \sim \sqrt{M^2/\la_1}$ is also a local minimum: away from that point the MSSM particles develop masses and again reduce $g_\textrm{eff}$. 

In fact, by counting relativistic degrees of freedom at temperatures $m_{\frac{3}{2}} \ll T \lesssim M$ one finds that $X_\phi$ is the global minimum. In the $h$-vacuum the relativistic species 
are the $\Phi, \Phib$ chiral multiplets and the U(1)$''$ gauge multiplet. In the $\phi$-vacuum, the particles of the MSSM are all light relative to $T$. Hence
\bea
f(0,T) &\simeq& - \frac{15}{2} \frac{\pi^2}{90}T^4, \\
f(X_\phi,T) &\simeq& -\frac{915}{4} \frac{\pi^2}{90}T^4.
 \eea
The minima of the free energy density are separated by a free energy barrier of height $\sim T^4$. 
The transition rate can be calculated in the standard way \cite{Linde:1978px} by calculating the free energy of
the critical bubble $E_c$, and it is not hard to show that the transition rate is suppressed by a factor $\exp(-X_\phi/T)$.  Hence we expect that the universe is trapped in the $h$-vacuum at temperatures $T \gtrsim m_X/q_\phi g'$.

\subsection{Gravitinos and dark matter}

Gravitinos are an inevitable consequence of supersymmetry and General Relativity, and there are strict constraints on their mass in the cosmological models with a standard thermal history and an R-symmetry guaranteeing the existence of a 
lightest supersymmetry particle (\lsp) \cite{Gravitino}.  
Even when unstable, they cause trouble either by decaying 
after nucleosynthesis and photodissociating light elements, or by decaying into the \lsp.  
The result is a constraint on the reheat temperature $T_\textrm{rh}$ in order to 
suppress the production of gravitinos.  
The relic abundance of thermally produced gravitinos is approximately 
\ben
Y_{\frac32} \simeq 2.4 \times 10^{-12} \om_{\tilde G} \left(\frac{T_\textrm{rh}}{10^{10}\; \textrm{GeV}}\right),
\een
where gravitinos are taken much more massive than the other superparticles, and $ \om_{\tilde G}$ is a factor taking into account the variation in the predictions. In recent literature it has taken the value 1.0 \cite{Rychkov:2007uq,Kawasaki:2008qe} and 0.6 \cite{BuchmullerEtAl}.
The \lsp\ density parameter arising from a particular relic abundance in the MSSM is  
\ben
\Omega_{\rm \lsp}h^2 \simeq 2.8\times 10^{10}\frac{m_{\rm \lsp}}{100\,\textrm{GeV}}Y_{\frac32}.
\een
The \lsp\ density
parameter from thermally produced gravitinos is therefore   
\ben \label{e:GraCon} \Omega_\textrm{\lsp}h^2
\simeq 6\times 10^{-2} \om_{\tilde G} \frac{m_\textrm{LSP}}{100\; \textrm{GeV}} 
\left(\frac{T_\textrm{rh}}{10^{10}\; \textrm{GeV}} \right)  ,
\een 
In our model, we will see that the gravitinos generated by the first stage of reheating, or by 
non-thermal production from decaying long-lived scalars~\cite{Endo:2006zj}, 
 are diluted by a period of thermal inflation. The constraint therefore applies to reheating after thermal inflation.

\subsection{Thermal inflation in the $h$-vacuum}
\label{s:TheInf}

In this section we continue with the assumption that the universe exits inflation into the $h$-vacuum.
As the temperature falls, eventually soft terms in the potential become
comparable to thermal energy density, and the universe can seek its true ground state, which 
we established in Section~\ref{s:HigPot} was $\chi = \pi/2$, the $\phi$-vacuum.  
This leads to a second period of inflation, akin to the complementary modular inflation model of \reference{Lazarides:2007dg}. Unlike this model, we
will see that reheating temperature is high enough to regenerate an interesting density of gravitinos, and also to allow baryogenesis by leptogenesis.

At zero temperature the difference in energy density between the $h$-vacuum and the
$\phi$-vacuum is (see \eqns{eq:phivac}{eq:hvac})
\ben
\De V_\textrm{eff}^0 \simeq s_c^2 \left(1 + 2\frac{\la_3^2}{\la_1^2} \right)^{\half}\left(\frac{m_\frac{3}{2}g'^2}{16\pi^2}\right)^2 4Qq_\phi^2\left(1 - \frac{q_{H_1}^2}{q_\phi^2}\frac{\la_1}{\la_3}\right). 
\een
Defining
an effective SUSY-breaking scale 
\ben
m_\textrm{sb} = \left(\frac{m_\frac{3}{2}g'^2}{16\pi^2}\right) q_\phi \sqrt{Q},
\een
we see that a period of thermal inflation \cite{Lyth:1995ka}
starts at 
\ben
T_\textrm{i} \simeq \left( \frac{30}{g_\textrm{eff}\pi^2}  s_c^2m_\textrm{sb}^2\right)^\frac14 .
\een
Using the CMB normalisation for $N$ e-foldings of standard hybrid inflation, $(s_c/\mpl) \simeq 3\times 10^{-3}$ (dropping the unimportant dependence on $N$), and 
the MSSM value for the degrees of freedom $g_\textrm{eff} = 915/4$, we have 
\ben
T_\textrm{i} \simeq 1.0 \times 10^{9} \left(\frac{m_\textrm{sb}}{1\;\textrm{TeV}}\right)^{\half} \; \textrm{GeV}.
\een
Thermal inflation continues until 
the quadratic term in the thermal potential $q_\phi^2{g'}^2T^2|X|^2$ becomes the same size
as the negative soft mass terms $m_X^2|X|^2$. 
%
Hence the transition which ends thermal inflation takes
place at $T_\textrm{e} \sim m_\textrm{sb}$, and the number of e-foldings
of thermal inflation is 
\ben
N_\th \simeq \half \ln \left( \frac{s_c}{m_\textrm{sb}}\right) \simeq 17, 
\een
taking $m_\textrm{sb} \sim 1$ TeV.  Thus any gravitinos will be diluted to unobservably low densities, as will any baryon number generated prior to thermal inflation.
 
There is another period of reheating as the energy of the modulus $X$ is converted to particles.  Around the true vacuum, the $X$ is mostly Higgs, and so its large amplitude oscillations will be quickly converted into the particles of the MSSM in much less than an expansion time, and the vacuum energy will be efficiently converted into thermal energy.  With the assumption of complete conversion of vacuum energy into thermal energy,  the reheat temperature following thermal inflation will be
\ben
T_\textrm{rh3} = \left( \frac{30}{g_\textrm{eff}\pi^2} \De V_\textrm{eff}^0 \right)^\frac14 \simeq T_\textrm{i}.
\een
This reheating regenerates the gravitinos, and we may again apply the gravitino constraint  \eqn{e:GraCon}, finding
\ben
\Omega_\textrm{\lsp}h^2
\simeq 6\times 10^{-3} \om_{\tilde G} \frac{m_\textrm{LSP}}{100\; \textrm{GeV}} 
\left(\frac{m_\textrm{sb}}{1\;\textrm{TeV}} \right)^{\half} .
\een
We can convert the relic density into a constraint on the gravitino mass, requiring that the \lsp\ density is less than or equal to the observed dark matter abundance, $\Om_\textrm{dm}h^2 \simeq 0.1$, obtaining
\ben
m_\frac32 \lesssim \frac{5\times 10^{4}}{{g'}^2q_\phi \sqrt{Q}} 
\left(\om_{\tilde G} \frac{m_\textrm{LSP}}{100\; \textrm{GeV}}\right)^{-2}\;\textrm{TeV}.
\een
Hence this class of models requires a high gravitino mass in order to saturate the bound and generate the dark matter.  

We can be a bit more precise if we use use the phenomenological relations derived in Section \ref{sec:spectrum}.  
Firstly, in order to fit $\mu_h$ we have from \eqn{e:MuCon}
\ben
q_\phi^2 {g'}^2 \simeq  \frac{\la_1}{\la_3},
\een
while we can derive a phenomenological formula for the \lsp\ mass from Table \ref{table:spectrumA}
\ben
m_\textrm{\lsp} \simeq 3.3\times 10^{-3} m_\frac32.
\een
Hence
\ben
m_\frac32 \lesssim {300} \left( \frac{1}{\om^2_{\tilde G}} \frac{q_\phi}{\sqrt{Q}}\frac{\la_3}{\la_1} \right)^{\frac{1}{3}}\;\textrm{TeV},
\een
with the inequality saturated if the gravitino decays supply all the dark matter. 

In the case where the dark matter consists of {\lsp}s derived from gravitino decay, we can derive a range of acceptable values for the gravitino mass, as we have a constraint (\ref{eq:la3range}) on $\frac{\la_3}{\la_1}$ from requiring the exit to a false $h$-vacuum. Hence, in order for gravitino-derived {\lsp}s in this model to comprise all the dark matter, we have
\ben
  \left(\frac{1}{\om^2_{\tilde G}} \frac{q_\phi}{\sqrt{Q}}\frac{q_{H_1}^2}{q_\phi^2} \right)^{\frac{1}{3}} 
\lesssim \frac{m_\frac32}{300\;\textrm{TeV}} \lesssim 
\left(\frac{1}{\om^2_{\tilde G}} \frac{q_\phi}{\sqrt{Q}} \right)^{\frac{1}{3}}.
\een
For example, taking $q_L=0$ as in Section \ref{sec:spectrum}, and recalling the range of the theoretical predictions $0.6 \lesssim \om_{\tilde G} \lesssim 1.0$, we find that $m_\frac{3}{2}$ is independent of $q_E$ and in the range 
\ben
{100}   \lesssim m_\frac32 \lesssim {430}  \;\textrm{TeV}.
\een
Interestingly, a Higgs with mass near 125 GeV also demands a high gravitino mass.  In order to fit the central value of $\delta a_\mu$ we require a gravitino mass of $60$ TeV, which would require $ \om_{\tilde G} \simeq 2$, or another source of dark matter.

\subsection{Cosmic string formation and constraints}

The breaking of the U(1)$''$ gauge symmetry at the end of thermal inflation results in the formation of cosmic strings 
\cite{Barreiro:1996dx,Hindmarsh:2011qj,Copeland:2011dx}. 
The string tension in models with flat directions is much less than the naive calculation, as the potential energy density in core the string is of order $\De V \sim s_c^2m^2_\textrm{sb}$ rather than $M^4$. 
The vacuum expectation of the modulus field defined in Section~\ref{sec:reheating}\ is still $X_0 \sim s_c$, so as a rough approximation we can therefore take the potential as
\ben
V_\textrm{string} \sim \frac{m^2_\textrm{sb}}{s_c^2}(X^2 - X_0^2)^2.
\een
showing that there is an effective scalar coupling of order $(m^2_\textrm{sb}/s_c^2)$. The string 
tension is approximately
\ben
\mu \simeq 2\pi \tenfun\left(\frac{m^2_\textrm{sb}}{q_\phi^2g'^2s_c^2}\right) \frac{2M^2}{\la_1},
\een 
where $\tenfun$ is a slowly varying function of its argument, with 
\cite{Hill:1987qx}
\ben
\tenfun(\beta) \simeq 
2.4/\ln(2/\beta), \; (\beta < 10^{-2}).
\een
Hence, for $q^2_\phi{g'}^2 = 2$, $s_c = 3\times 10^{-3}\mpl$, and $m_\textrm{sb} = 1$ TeV as above,
\ben
\tenfun \simeq 0.04,
\een
demonstrating that the string tension is more than an order of magnitude below its naive value $4\pi s_c^2$, which reduces the CMB constraint on this model.
Hence the string tension in this model is 
\ben
G\mu = \frac{B}{4} \frac{s_c^2}{\mpl^2} \sim 10^{-7},
\een
well below the 95\% confidence limit for CMB fluctuations from strings
\cite{Dunkley:2010ge,Urrestilla:2011gr}. 

There are also other bounds on strings depending on uncertain details
about their primary decay channel.  Pulsar timing provides a strong
bound if the long strings lose a significant proportion of energy into
loops with sizes above a light year or so (smaller loops radiate at
frequencies to which pulsar timing is not very sensitive). In this case
recent European Pulsar Timing Array data \cite{vanHaasteren:2011ni} can
be used to place a conservative upper bound of $G\mu < 5.3 \times
10^{-7}$ \cite{Sanidas:2012ee} for strings with a reconnection
probability of close to unity (as is the case in field theory), and
loops formed with a typical size of about $10^{-5}$ of the horizon size.
Future experiments will place tighter (but still model-dependent) bounds
\cite{Sanidas:2012ee,Kuroyanagi:2012wm}.  For example, the Large
European Array for Pulsars (LEAP) will be two orders of magnitude more
sensitive than EPTA \cite{LEAP} and will be able to detect the
gravitational radiation from the loops in this model if they are large
enough to radiate into the LEAP sensitivity window.  Current string
modelling \cite{BlancoPillado:2011dq} indicates this is likely if loop
production is significant.

Strings may also produce high energy particles, whose decays can produce
cosmic rays over a very wide spectrum of energies. If $f_\textrm{cr}$ is
the fraction of the energy density going into cosmic rays, then the 
diffuse $\gamma$-ray background provides a limit \cite{Hindmarsh:2011qj}
$
G\mu \lesssim 10^{-10} f^{-1}_\textrm{cr}.
$
Given that the strings in our model contain a large Higgs condensate, we
would expect that all particles produced by the strings would end up as
Standard Model particles or neutralinos. Thus we require that the decays
are primarily gravitational in order to avoid the cosmic ray bound.

\subsection{Baryogenesis}

Baryon asymmetry requires baryon number (B) violation, C violation, and
CP violation \cite{Sakharov:1967dj}.    In common with the standard
model, our model has C violation and sphaleron-induced B violation.  It
can also support  CP-violating phases in the neutrino Yukawa couplings. 
In \cite{Basboll:2011mh}, it was pointed out that leptogenesis
\cite{Fukugita:1986hr} was natural in the model, provided that the reheat temperature is greater
than about $10^9$ GeV.  

As we saw in Section \ref{s:TheInf}, this is the approximate value of the reheat temperature 
after thermal inflation, and so we require at least one right-handed neutrino which is sufficiently light to be generated 
in the reheating process, i.e. with a mass less than around $10^9$ GeV.  The baryogenesis in our model should therefore be similar to 
that of \reference{Buchmuller:2012wn}.  

In \samsb, the light scalars are weakly coupled to the Higgs (the stops
are both at the TeV scale), and  so the electroweak phase transition is 
a crossover \cite{Laine:1998qk}.  This means that there is no
conventional electroweak baryogenesis (see e.g.\
\reference{Schmidt:2011zza} for a recent review).

\section{Conclusions}

The \samsb\ model, as described here, is in our opinion the most
attractive way of resolving  the tachyonic slepton problem of anomaly
mediated supersymmetry  breaking. The low energy spectrum is similar to
that of regions of \cmssm\ or \msugra\  parameter space,  but with
characteristic features, most notably a wino \lsp.  We have seen that,
while it is possible to obtain a light \sm-like Higgs with a mass of 
125\GeV, this requires fine-tuning and also results in a suppression  of
the \sic\ contribution to $a_{\mu}$,  so that the current theoretical
prediction  for $a_{\mu}$ in our model is about $3\sigma$ below the
experimental value.\footnote{This tension has also been noted in the \cmssm\  
\cite{Buchmueller:2011sw}, underlining the importance of an 
independent experimental measurement.}


Moreover, to produce a Higgs of over 120 \GeV, we need to increase the
gravitino mass to over 80 \TeV.  If the gravitino mass is over 100 TeV
we can use wino {\lsp}s derived from gravitino decays to account for all
the dark matter.   

Assuming that the $\textrm{U(1)}'$ introduced to solve  the tachyonic
slepton problem is broken at a high scale, $M$,  we have seen that 
\samsb\ naturally realises F-term hybrid inflation. 
The universe may exit the inflationary era into a vacuum dominated by 
large vevs for the \mssm\ Higgs fields, $h_{1,2}$, with the  true vacuum
with unbroken $\textrm{SU(3)}\otimes \textrm{SU(2)} \otimes
\textrm{U(1)}_Y$ (above the electroweak scale)  attained only after a
later period of approximately 17 e-foldings of thermal inflation.

The thermal inflation reduces the number of e-foldings of high-scale
inflation to about 40, and hence   the spectral index of scalar CMB
fluctuations is reduced to about 0.975, within about $1\sigma$ of the
WMAP7 value.  Cosmic strings are formed at the end of thermal inflation,
with a low mass per unit length, satisfying observational bounds
provided  their main decay channel is gravitational, and the typical
size of string loops at formation is about $10^{-5}$ of the horizon
size, or so small that they radiate at a frequency below 1 yr$^{-1}$, to
which pulsar timing is not sensitive.  The Large European Array for
Pulsars will be two orders of magnitude more sensitive, and be capable
of closing the window in the loop size at $10^{-5}$ of the horizon, or
detecting the gravitational radiation.

\section*{\large Acknowledgements}

This research was supported in part by the Science and Technology
Research Council [grant numbers ST/J000477/1 and ST/J000493/1]. 
Part of it was done one of us (DRTJ) was visiting the Aspen Center for
Physics.

\end{document}